\RequirePackage{fixltx2e}
\documentclass[a4paper,fleqn,reqno,english,journal=cmatex,manuscript=article]{revtex4-2}
\usepackage[LGR,T1]{fontenc}
\usepackage[latin9]{inputenc}
\usepackage{babel}
\usepackage{array}
\usepackage{float}
\usepackage{textcomp}
\usepackage{multirow}
\usepackage{amsmath}
\usepackage{amsthm}
\usepackage{amssymb}
\usepackage{graphicx}
\usepackage[unicode=true,pdfusetitle,
 bookmarks=true,bookmarksnumbered=false,bookmarksopen=false,
 breaklinks=false,pdfborder={0 0 1},backref=false,colorlinks=false]
 {hyperref}

\makeatletter

\DeclareRobustCommand{\greektext}{%
  \fontencoding{LGR}\selectfont\def\encodingdefault{LGR}}
\DeclareRobustCommand{\textgreek}[1]{\leavevmode{\greektext #1}}
\ProvideTextCommand{\~}{LGR}[1]{\char126#1}

\newcommand{\lyxmathsym}[1]{\ifmmode\begingroup\def\b@ld{bold}
  \text{\ifx\math@version\b@ld\bfseries\fi#1}\endgroup\else#1\fi}

\providecommand{\tabularnewline}{\\}
\newcommand{\lyxdot}{.}

\usepackage{lmodern}

\makeatother

\begin{document}
\title{Optimizing Lead-Free Chalcogenide Perovskites for High-Efficiency
Photovoltaics via Alloying Strategies}
\author{Surajit Adhikari{*}, Priya Johari}
\email{sa731@snu.edu.in, priya.johari@snu.edu.in}

\affiliation{Department of Physics, School of Natural Sciences, Shiv Nadar Institution
of Eminence, Greater Noida, Gautam Buddha Nagar, Uttar Pradesh 201314,
India}
\begin{abstract}
Lead-free chalcogenide perovskites are emerging as game-changers in
the race for sustainable, high-performance photovoltaics. These materials
offer a perfect trifecta: non-toxic elemental composition, exceptional
phase stability, and outstanding optoelectronic properties. However,
unlocking their full potential for solar cell applications requires
advanced strategies to fine-tune their electronic and optical behavior.
In this study, we take CaHfS$_{3}$\textemdash a promising but underexplored
candidate\textemdash and revolutionize its performance by introducing
targeted substitutions: Ti at the cation site and Se at the anion
site. Using cutting-edge computational techniques, including density
functional theory, GW calculations, and the Bethe-Salpeter equation
(BSE), we reveal how these substitutions transform the material's
properties. Our findings highlight that alloyed compounds such as
CaHfS$_{3-x}$Se$_{x}$ and CaHf$_{1-y}$Ti$_{y}$X$_{3}$ (X = S,
Se) are not only phase-stable but also feature adjustable direct G$_{0}$W$_{0}$@PBE
bandgaps (1.29\textendash 2.67 eV), reduced exciton binding energies,
and significantly improved polaron mobility. These modifications enable
better light absorption, reduced electron-hole recombination, longer
exciton lifetimes, and enhanced quantum yield. Impressively, the alloyed
perovskites, specifically, for the Ti-rich Se-based perovskites, achieve
a spectroscopic-limited maximum efficiency of up to 28.06\%, outperforming
traditional lead-based halide perovskites. Our results demonstrate
that strategic alloying is a powerful tool to supercharge the optoelectronic
properties of lead-free chalcogenide perovskites, positioning them
as strong contenders for next-generation photovoltaic technologies.
\end{abstract}
\maketitle

\section{Introduction:}

Inorganic\textminus organic halide perovskites (IOHPs) have garnered
unparalleled attention over the past decade due to their exceptional
electronic and optical properties \citep{chapter2-12,chapter3-8,chapter3-9,chapter3-10}.
Consequently, the popularity of IOHPs in solar cell research has skyrocketed
in recent years, with their record power conversion efficiency (PCE)
increasing dramatically from 3.8\% to 26.7\% \citep{chapter2-12,chapter2-53}.
Despite these advances, a critical challenge persists: the majority
of IOHPs are lead (Pb)-based, raising serious concerns about toxicity.
Additionally, issues such as degradability, along with thermal and
chemical instability attributed to the organic components, present
formidable barriers to their large-scale industrial deployment \citep{chapter3-12,chapter1-16}.
These challenges have driven research toward designing stable, nontoxic
alternative perovskites, paving the way for new, efficient photovoltaic
materials.

Recently, chalcogenide perovskites (CPs) materials have garnered significant
attention as promising alternatives to lead-based perovskites for
solar cell technology \citep{chapter3-11,chapter3-16,chapter3-17,chapter3-18}.
The chemical formula of CPs is represented as ABX$_{3}$ \citep{chapter3-3,chapter3-18},
where A and B correspond to divalent alkali-earth metal cations (Ca$^{2+}$,
Sr$^{2+}$, Ba$^{2+}$) and tetravalent transition metal cations (Ti$^{4+}$,
Zr$^{4+}$, Hf$^{4+}$), respectively. The X site is typically occupied
by a chalcogen anion such as S$^{2-}$ or Se$^{2-}$. Numerous experimental
and theoretical studies have successfully demonstrated the synthesis
of CPs, revealing their fascinating and promising properties \citep{chapter3-11,chapter3-13,chapter3-16,chapter3-17,chapter3-18,chapter3-19,chapter3-37}.
Also, at ambient temperature, two distinct phases of CPs have been
identified: a needle-like orthorhombic phase (NH$_{4}$CdCl$_{3}$-type)
and a distorted orthorhombic structure (GdFeO$_{3}$-type). Despite
their structural differences, both phases share the same space group,
$Pnma$ (No. 62) \citep{chapter3-18}.

Among these CPs, ABS$_{3}$ (A = Ca, Sr, Ba; B = Zr, Hf) have been
experimentally synthesized with distorted structures featuring corner-sharing
BS$_{6}$ octahedra connected in a three-dimensional network \citep{chapter3-37}.
Also, a distorted CaTiS$_{3}$ CP has been recently synthesized experimentally,
exhibiting a bandgap of 1.59 eV \citep{chapter6-1}. In addition,
several first-principles-based studies have predicted that ABX$_{3}$
(A = Ca, Sr, Ba; B = Zr, Hf and X = S, Se) CPs exhibit promising electronic
and optical properties, making them attractive for optoelectronic
applications \citep{chapter3-19,chapter1-63,chapter5-16}. However,
the performance of ABS$_{3}$ (A = Ca, Sr, Ba; B = Zr, Hf) CPs in
photovoltaic applications is significantly hindered compared to conventional
lead-based HPs \citep{chapter5-11,chapter5-12,chapter5-14,chapter5-13,chapter5-15,chapter5-17,chapter2-20},
due to their higher bandgaps (1.83$-$2.46 eV), large exciton binding
energies (0.19$-$0.26 eV), lower charge carrier mobilities (6.84$-$18.77
cm$^{2}$V$^{-1}$s$^{-1}$), and reduced PCE (10.56$-$25.02\%) \citep{chapter3-19,chapter1-63,chapter3-37}.
Another study shows that selenide-based CPs outperform sulfur-based
CPs in photovoltaic properties \citep{chapter5-16}. However, their
performance remains inferior to that of lead-based conventional HPs,
with lower charge carrier mobilities (56.08$-$77.59 cm$^{2}$V$^{-1}$s$^{-1}$)
and reduced PCE (17.45$-$23.08\%). Therefore, it is essential to
investigate CPs with optimal photovoltaic properties to address these
challenges effectively.

It is well-known that the BX$_{6}$ octahedra, which form the core
of ABX$_{3}$ perovskite materials, play a crucial role in determining
the material's electronic band structure, influencing properties such
as the bandgap, charge carrier mobility, and efficiency. Recently,
several studies have focused on tuning the electronic structure of
CPs by chemically substituting cations or anions at the B and X sites.
For instance, Meng et al. demonstrated a significant reduction in
the bandgap of BaZrS$_{3}$ through first-principles simulations by
alloying Ti with Zr. They reported that doping the perovskite (BaZr$_{1-x}$Ti$_{x}$S$_{3}$,
$x$ = 0.1) reduced the bandgap to 1.47 eV, making it suitable for
single-junction solar cell applications \citep{chapter6-4}. Similarly,
several experimental and first-principles-based studies have shown
that alloying in CPs, such as BaHf$_{1-x}$Ti$_{x}$S$_{3}$ (0 $\leq$
$x$ $\leq$ 0.05) \citep{chapter6-3}, BaHf$_{1-x}$Zr$_{x}$S$_{3}$
($x$ $\leq$ 1) \citep{chapter6-6}, and BaZr$_{1-x}$Sn$_{x}$S$_{3}$
($x$ $\leq$ 0.25) \citep{chapter6-5}, significantly enhances their
photovoltaic properties. Liu et al. also explored anion-site engineering
in Hf-based CPs to tailor their optoelectronic properties and enhance
their photovoltaic performance \citep{chapter6-7}. However, the excitonic
and polaronic properties of the above mentioned mixed and alloyed
CPs have not been explored for efficient solar cell applications,
largely due to the substantial computational demands.

Inspired by the alloying strategies in CPs for designing efficient
materials for optoelectronic applications, this study takes a systematic
leap into designing next-gen materials for high-performance optoelectronic
applications. We delve into the comprehensive study on the electronic,
optical, transport, excitonic, and polaronic properties of CaHfS$_{3-x}$Se$_{x}$
($x$ = 0, 1, 2, 3) and CaHf$_{1-y}$Ti$_{y}$X$_{3}$ ($y$ = 0,
0.25, 0.50, 0.75, 1; X = S and Se) compounds within the framework
of density functional theory (DFT) \citep{chapter2-36,chapter2-37}
and many-body perturbation theory (MBPT) \citep{chapter3-1,chapter3-2}.
Our findings are electrifying: these CPs exhibit direct G$_{0}$W$_{0}$@PBE
\citep{chapter1-69,chapter1-70} bandgaps in the ideal range of 1.29\textendash 2.67
eV, setting the stage for exceptional optoelectronic performance.
Leveraging the Bethe-Salpeter equation (BSE) method \citep{chapter1-67,chapter1-68},
we calculated exciton binding energies ($E_{B}$) between 0.008 and
0.204 eV, ensuring robust excitonic stability. Adding depth, we employed
density functional perturbation theory (DFPT) \citep{chapter1-60}
to assess ionic contributions to dielectric behavior, followed by
the Fr\"ohlich and Hellwarth models to explore polaronic phenomena.
Our findings indicate that these CPs exhibit carrier-phonon coupling
paired with remarkably high polaron mobility, far outpacing traditional
lead-based CPs. Pushing boundaries further, we evaluated the spectroscopic
limited maximum efficiency (SLME), revealing an impressive potential
efficiency of up to 28.06\%\textemdash a breakthrough for solar cell
technology. These results underscore the power of alloying as a game-changing
strategy to fine-tune the optoelectronic properties of CPs, paving
the way for more efficient, sustainable, and lead-free alternatives
in energy applications. This work highlights the transformative potential
of alloying to unlock high-efficiency materials and sets a new benchmark
for the future of optoelectronics.

\section{Computational Details:}

In this paper, state-of-the-art calculations based on first-principles
density functional theory (DFT) \citep{chapter2-36,chapter2-37},
density functional perturbation theory (DFPT) \citep{chapter1-60},
and many-body perturbation theory (MBPT) \citep{chapter3-1,chapter3-2}
were conducted using the Vienna Ab initio Simulation Package (VASP)
\citep{chapter1-31,chapter1-32}. The interactions between the valence
electrons and the ion cores were described using projector augmented
wave (PAW) pseudopotentials \citep{chapter1-33}. The PAW potentials
with valence-electron configurations considered for Ca, Ti, Hf, S,
and Se were 3s$^{2}$3p$^{6}$4s$^{2}$, 3p$^{6}$4s$^{2}$3d$^{2}$,
5p$^{6}$6s$^{2}$5d$^{2}$, 3s$^{2}$3p$^{4}$, and 4s$^{2}$4p$^{4}$,
respectively. The structural optimizations were performed using the
Perdew-Burke-Ernzerhof (PBE) exchange-correlation (xc) functional
within the generalized gradient approximation (GGA) \citep{chapter1-34},
which accounts for electron-electron interactions. The kinetic energy
cutoff was set to 400 eV and the electronic self-consistent-field
iteration energy convergence criteria was chosen at 10$^{-6}$ eV.
The lattice parameters as well as coordinates of all the atoms are
fully optimized until the Helmann-Feynman forces on each atom's were
less than 0.01 eV/$\textrm{\AA}$. In order to determine the optimized structures,
Brillouin zone integrations were performed using a $7\times7\times5$
$\Gamma$-centered $\mathbf{k}$-point grid. The optimized crystal
structures were visualized using the VESTA (Visualization for Electronic
and STructural Analysis) software \citep{chapter2-3}. The phonon
dispersion curves were computed using the DFPT method with $2\times2\times2$
supercells, employing the PHONOPY package \citep{chapter3-6}. The
electronic band structures were computed using the PBE xc functional,
incorporating the spin-orbit coupling (SOC) effect, although it does
not influence the overall trend. In addition, the electronic bandgaps
were simultaneously calculated using the hybrid HSE06 xc functional
\citep{chapter1-35} and the MBPT-based G$_{0}$W$_{0}$ method \citep{chapter1-69,chapter1-70}
for more accurate estimation. The initial step of the G$_{0}$W$_{0}$
calculation was carried out using the PBE xc functional. The effective
masses of the charge carriers were estimated using the SUMO code,
which performed a parabolic fitting at the band edges. The Bethe-Salpeter
equation (BSE) \citep{chapter1-67,chapter1-68} calculations were
also performed on top of the single-shot GW (G$_{0}$W$_{0}$)@PBE
to more accurately estimate the optical properties, specifically incorporating
the electron-hole interaction. Here, a $\Gamma$-centered $3\times3\times2$
$\mathit{\mathbf{k}}$-grid and a converged number of bands (NBANDS)
of 640 were used for the GW-BSE calculations. The electron-hole kernel
for the BSE calculations was generated using 24 occupied and 24 unoccupied
bands. The VASPKIT \citep{chapter1-48} package was employed for the
post-processing of the elastic and optical properties. The ionic contribution
to the dielectric constant was also determined using the DFPT method.

Using the hydrogenic Wannier-Mott (WM) model, the exciton binding
energy ($E_{B}$) for an electron-hole pair with screened Coulomb
interaction is calculated as follows \citep{chapter2-38,chapter5-16}:
\begin{equation}
E_{B}=\Big(\frac{\mu^{\ast}}{m_{0}\varepsilon_{\mathrm{eff}}^{2}}\Big)R_{\infty}\label{eq:1}
\end{equation}
where $\mu^{\ast}$ is the reduced mass of the charge carriers, $m_{0}$
represents the rest mass of electron, $\varepsilon_{\mathrm{eff}}$
denotes the effective dielectric constant, and $R_{\infty}$ is the
Rydberg constant. The effective mass has been determined using the
following equation \citep{chapter3-35}:

\begin{equation}
m^{*}=3\left[\frac{1}{m_{xx}^{*}}+\frac{1}{m_{yy}^{*}}+\frac{1}{m_{zz}^{*}}\right]^{-1}\label{eq:2}
\end{equation}

where $m_{ii}^{\ast}$ is the effective mass in the $i$-th direction
(i = x, y, z). The reduced mass of the carrier, $\mu^{\ast}$, is
given by,
\begin{equation}
\mathrm{\frac{1}{\mu^{*}}=\frac{1}{\mathit{m_{e}^{*}}}+\frac{1}{\mathit{m_{h}^{*}}}}.\label{eq:3}
\end{equation}

The correction to the exciton binding energy ($E_{B}$) due to phonon
screening is expressed as \citep{chapter3-39}:

\begin{equation}
\Delta E_{B}^{ph}=-2\omega_{LO}\left(1-\frac{\varepsilon_{\infty}}{\varepsilon_{static}}\right)\frac{\sqrt{1+\omega_{LO}/E_{B}}+3}{\left(1+\sqrt{1+\omega_{LO}/E_{B}}\right)^{3}},\label{eq:4}
\end{equation}

where $\omega_{LO}$ denotes the characteristic phonon angular frequency,
and $\varepsilon_{\infty}$ and $\varepsilon_{static}$ represent
the electronic (optical) and static (electronic + ionic) dielectric
constants, respectively. The thermal \textquotedbl B\textquotedbl{}
approach developed by Hellwarth et al. \citep{chapter2-22} is employed
to determine $\omega_{LO}$ by averaging the spectral contributions
of multiple phonon branches (for details, see the Supplemental Material).

The exciton radius ($r_{exc}$) is computed using the following formula
\citep{chapter2-38,chapter5-16}:

\begin{equation}
r_{exc}=\frac{m_{0}}{\mu^{*}}\mathrm{\varepsilon_{eff}}n^{2}r_{Ry},\label{eq:5}
\end{equation}

where $n$ represents the exciton energy level and $r_{Ry}$ = 0.0529
nm is the Bohr radius. In our study, the electronic contribution to
the dielectric function ($\varepsilon_{\infty}$) is taken as the
effective value and $n=1$, which yields the smallest exciton radius.
Using the exciton radius, the probability of the wave function at
zero separation ($|\phi_{n}(0)|^{2}$) can be calculated as follows
\citep{chapter5-16,chapter5-18}:

\begin{equation}
|\phi_{n}(0)|^{2}=\frac{1}{\pi(r_{exc})^{3}n^{3}}.\label{eq:6}
\end{equation}

Within the framework of Fr\"ohlich's polaron model, the interaction
between longitudinal optical phonons and an electron traveling through
the lattice is governed by the dimensionless Fr\"ohlich parameter, $\alpha$,
which is expressed as \citep{chapter5-16},
\begin{equation}
\alpha=\frac{1}{4\pi\varepsilon_{0}}\frac{1}{2}\Big(\frac{1}{\varepsilon_{\infty}}-\frac{1}{\varepsilon_{static}}\Big)\frac{e^{2}}{\hbar\omega_{LO}}\Big(\frac{2m^{\ast}\omega_{LO}}{\hbar}\Big)^{1/2}\label{eq:7}
\end{equation}
where $\varepsilon_{0}$ represents the permittivity of free space
and $m^{\ast}$ denotes the effective mass of the carrier. By determining
the value of $\alpha$, the polaron energy ($E_{p}$) can be calculated
using the following expression \citep{chapter5-16}:
\begin{equation}
E_{p}=(-\alpha-0.0123\alpha^{2})\hbar\omega_{LO}.\label{eq:8}
\end{equation}
Using Feynman\textquoteright s approach, the polaron\textquoteright s
effective mass ($m_{p}$) can be expressed as follows (for a small
$\alpha$) \citep{chapter2-23}:
\begin{equation}
m_{p}=m^{\ast}\Big(1+\frac{\alpha}{6}+\frac{\alpha^{2}}{40}+...\Big)\label{eq:9}
\end{equation}
Finally, the polaron mobility ($\mu_{p}$) is estimated using the
Hellwarth polaron model \citep{chapter2-22} as follows:
\begin{equation}
\mu_{p}=\frac{\left(3\sqrt{\pi}e\right)}{2\pi c\omega_{LO}m^{*}\alpha}\frac{\sinh(\beta/2)}{\beta^{5/2}}\frac{w^{3}}{v^{3}}\frac{1}{K(a,b)}\label{eq:10}
\end{equation}
where $e$ denontes the electronic charge, $\beta=hc\omega_{LO}/k_{B}T$,
$w$ and $v$ both are the temperature-dependent variational parameters,
and $K(a,b)$ is a function of $\beta$, $w$, and $v$ (for details,
see the Supplemental Material).

\section{Results and Discussions:}

In this present study, we conducted a comprehensive investigation
to evaluate the effects of doping in both the cationic and anionic
sites on the optoelectronic properties of CaHfS$_{3}$ chalcogenide
perovskite. The following sections provide an in-depth analysis of
the stability, structural and electronic properties, transport behavior,
optical properties, excitonic dynamics, polaronic effects, and the
spectroscopic limited maximum efficiency (SLME) of CaHfS$_{3-x}$Se$_{x}$
($x$ = 0, 1, 2, 3) and CaHf$_{1-y}$Ti$_{y}$X$_{3}$ ($y$ = 0,
0.25, 0.50, 0.75, 1; X = S and Se) chalcogenide perovskites. These
findings aim to build a fundamental understanding and provide valuable
insights to guide future experimental investigations.

\begin{figure}[H]
\begin{centering}
\includegraphics[width=1\textwidth,height=1\textheight,keepaspectratio]{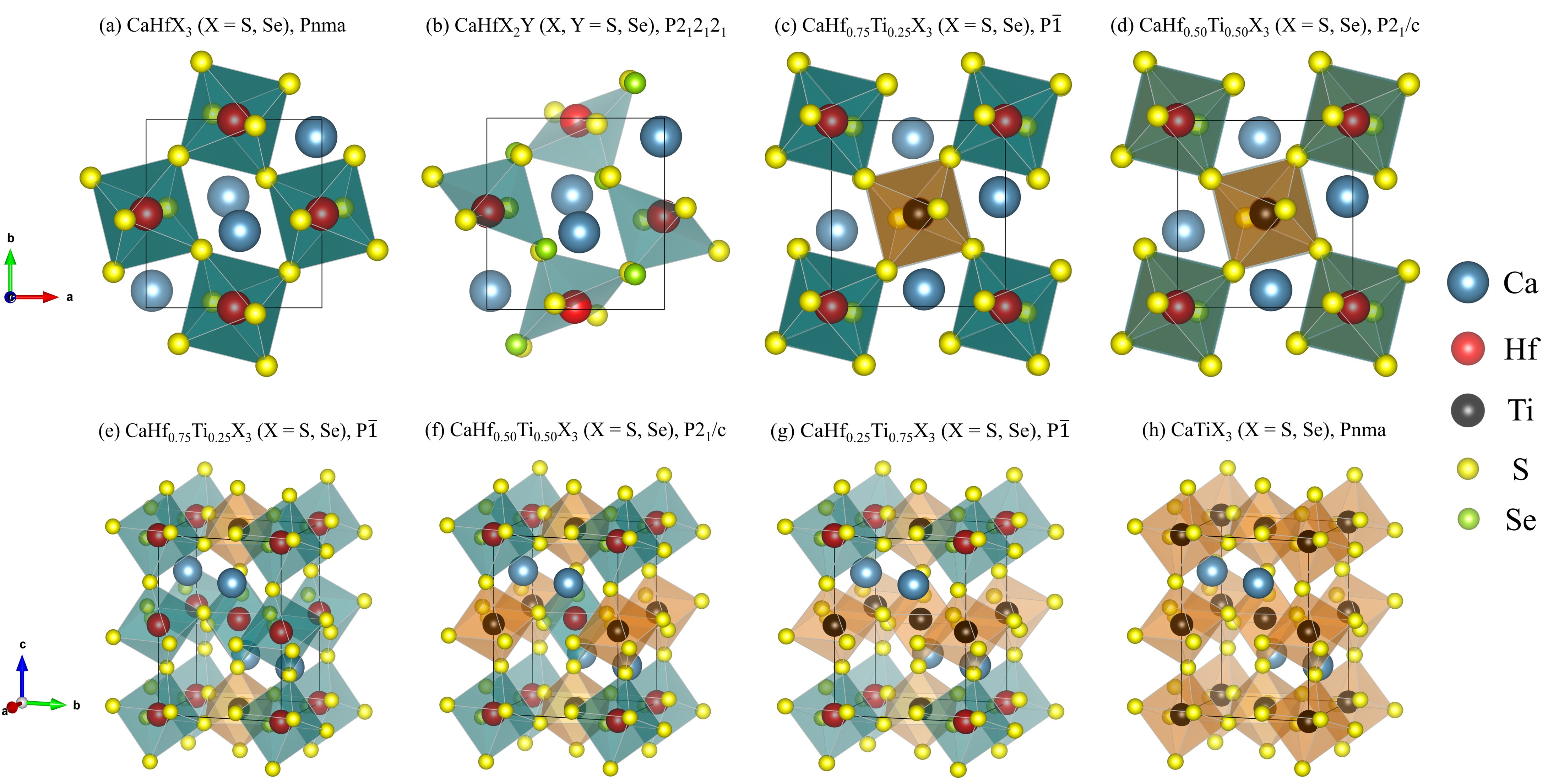}
\par\end{centering}
\centering{}\caption{\label{fig:1}Different view of the crystal structures of (a) CaHfX$_{3}$
(X=S, Se), (b) CaHfX$_{2}$Y/XY$_{2}$ (X, Y=S, Se; X$\protect\neq$Y),
(c, e) CaHf$_{0.75}$Ti$_{0.25}$X$_{3}$ (X=S, Se), (d, f) CaHf$_{0.50}$Ti$_{0.50}$X$_{3}$
(X=S, Se), (g) CaHf$_{0.25}$Ti$_{0.75}$X$_{3}$ (X=S, Se), and (h)
CaTiX$_{3}$ (X=S, Se) CPs, respectively.}
\end{figure}

\subsection{Structural Properties:}

\subsubsection{Crystal structure and crystallographic stability:}

Figure \ref{fig:1}(a) illustrates the crystal structure of the distorted
orthorhombic phase of the chalcogenide perovskite CaHfX$_{3}$ (X
= S, Se), which belongs to the space group $Pnma$ (No. 62). The crystal
structures of these compounds typically consist of four unit cells
containing a total of 20 atoms, including 4 Ca, 4 Hf, and 12 S/Se
atoms. In the distorted phase, Ca cation exhibits 12-fold coordination,
forming cuboctahedral structures with chalcogenides X (S or Se). Meanwhile,
the Hf cation undergoes 6-fold coordination with chalcogen atoms X
(S or Se), resulting in the formation of corner-sharing {[}HfX$_{6}${]}$^{8-}$
distorted octahedra, characterized by tilting and structural distortion.
To form mixed CP CaHfS$_{3-x}$Se$_{x}$, sulfur atoms are progressively
substituted by selenium atoms. Similarly, in the alloy CaHf$_{1-y}$Ti$_{y}$X$_{3}$,
hafnium atoms are partially replaced by titanium atoms. The crystal
structure of CaHfS$_{3-x}$Se$_{x}$ ($x$ = 1, 2) transitions to
a lower-symmetry orthorhombic structure {[}space group $P2_{1}2_{1}2_{1}$
(19){]} when one chalcogenide site is replaced by another. Similarly,
the crystal structures of CaHf$_{1-y}$Ti$_{y}$X$_{3}$ alloys undergo
symmetry reduction when hafnium is partially substituted by titanium.
For $y$ = 0.25 or 0.75, the structures reduce to a triclinic phase
{[}space group $P\bar{1}$ (2){]}, while for $y$ = 0.50, they adopt
a monoclinic phase {[}space group $P2_{1}/c$ (14){]}. When $y$ =
1, where hafnium is fully replaced by titanium, the structure transitions
back to a higher-symmetry orthorhombic phase {[}space group $Pnma$
(62){]}. These systems are modeled using their primitive cell, comprising
20 atoms. The primitive crystal structures of CaHfS$_{3-x}$Se$_{x}$
($x$ = 1, 2) and CaHf$_{1-y}$Ti$_{y}$X$_{3}$ ($y$ = 0.25, 0.5,
0.75, 1; X = S, Se) CPs are also presented in Figure \ref{fig:1}(b)-(h).

The calculated lattice parameters, space group (no.), and volume of
the mixed and alloyed CPs are presented in Table \ref{tab:1}. Notably,
the computed lattice parameters for CaHfS$_{3}$ show excellent agreement
with the available experimental result (a = 6.52 $\textrm{\AA}$, b
= 6.98 $\textrm{\AA}$, c = 9.54 $\textrm{\AA}$) \citep{chapter3-37}.
It is also observed that the lattice parameters and volume of these
mixed CPs increase with higher selenium content, attributed to the
larger ionic radius of Se (1.98 $\textrm{\AA}$) compared to S (1.84
$\textrm{\AA}$). Conversely, in alloyed CPs with higher titanium content,
the lattice parameters and volume decrease due to the smaller ionic
radius of Ti (0.605 $\textrm{\AA}$) compared to Hf (0.71 $\textrm{\AA}$).
In addition, the octahedral distortion parameters, including the average
bond length, bond angle variance, polyhedral volume, and distortion
index (bond length) for the HfX$_{6}$ and TiX$_{6}$octahedra in
CaHf$_{1-y}$Ti$_{y}$X$_{3}$ ($y$ = 0, 0.25, 0.5, 0.75, 1; X =
S, Se) CPs, are computed and presented in Table S1. A very low distortion
index for the HfX$_{6}$ and TiX$_{6}$ octahedra in these CPs suggests
that the octahedra remain nearly ideal across the composition range,
leading to enhanced structural stability, uniform electronic behavior,
and predictable optoelectronic properties. This is beneficial for
applications where consistent and tunable properties are essential,
such as optoelectronic devices.

\begin{table}[H]
\caption{\label{tab:1}Calculated lattice parameters, space group (no.), volume,
and decomposition energy of CaHfS$_{3-x}$Se$_{x}$ ($x$ = 0, 1,
2, 3) and CaHf$_{1-y}$Ti$_{y}$X$_{3}$ ($y$ = 0, 0.25, 0.5, 0.75,
1; X = S, Se) CPs.}

\centering{}{\scriptsize{}}%
\begin{tabular}{cccccccccc}
\hline 
\multirow{2}{*}{{\scriptsize{}Configurations}} &  & \multicolumn{3}{c}{{\scriptsize{}Lattice Parameters ($\textrm{\AA}$)}} &  & {\scriptsize{}Space group} & {\scriptsize{}Volume} & {\scriptsize{}$\mathrm{\mathit{E}_{d}(B)}$} & {\scriptsize{}$\mathrm{\mathit{E}_{d}(T)}$}\tabularnewline
\cline{3-5} \cline{4-5} \cline{5-5} 
 &  & {\scriptsize{}a} & {\scriptsize{}b} & {\scriptsize{}c} &  & {\scriptsize{}(No.)} & {\scriptsize{}($\textrm{\AA}^{3}$)} & {\scriptsize{}(eV/atom)} & {\scriptsize{}(eV/atom)}\tabularnewline
\hline 
\multicolumn{1}{c}{{\scriptsize{}CaHfS$_{3}$}} &  & {\scriptsize{}6.56} & {\scriptsize{}7.01} & {\scriptsize{}9.58} &  & {\scriptsize{}$Pnma$ (62)} & {\scriptsize{}440.87} & {\scriptsize{}-0.0365} & {\scriptsize{}-}\tabularnewline
{\scriptsize{}CaHfS$_{2}$Se} &  & {\scriptsize{}6.68} & {\scriptsize{}7.19} & {\scriptsize{}9.61} &  & {\scriptsize{}$P2_{1}2_{1}2_{1}$ (19)} & {\scriptsize{}460.96} & {\scriptsize{}-0.0617} & {\scriptsize{}0.0001}\tabularnewline
{\scriptsize{}CaHfSSe$_{2}$} &  & {\scriptsize{}6.79} & {\scriptsize{}7.28} & {\scriptsize{}9.74} &  & {\scriptsize{}$P2_{1}2_{1}2_{1}$ (19)} & {\scriptsize{}481.45} & {\scriptsize{}-0.0423} & {\scriptsize{}0.0063}\tabularnewline
{\scriptsize{}CaHfSe$_{3}$} &  & {\scriptsize{}6.84} & {\scriptsize{}7.34} & {\scriptsize{}10.03} &  & {\scriptsize{}$Pnma$ (62)} & {\scriptsize{}504.27} & {\scriptsize{}-0.0739} & {\scriptsize{}-}\tabularnewline
\multicolumn{1}{c}{{\scriptsize{}CaHf$_{0.75}$Ti$_{0.25}$S$_{3}$}} &  & {\scriptsize{}6.51} & {\scriptsize{}6.96} & {\scriptsize{}9.51} &  & {\scriptsize{}$P\bar{1}$ (2)} & {\scriptsize{}431.49} & {\scriptsize{}-0.0491} & {\scriptsize{}-0.0048}\tabularnewline
{\scriptsize{}CaHf$_{0.75}$Ti$_{0.25}$Se$_{3}$} &  & {\scriptsize{}6.79} & {\scriptsize{}7.29} & {\scriptsize{}9.96} &  & {\scriptsize{}$P\bar{1}$ (2)} & {\scriptsize{}493.72} & {\scriptsize{}-0.0846} & {\scriptsize{}-0.0039}\tabularnewline
\multicolumn{1}{c}{{\scriptsize{}CaHf$_{0.50}$Ti$_{0.50}$S$_{3}$}} &  & {\scriptsize{}6.47} & {\scriptsize{}6.90} & {\scriptsize{}9.43} &  & {\scriptsize{}$P2_{1}/c$ (14)} & {\scriptsize{}421.37} & {\scriptsize{}-0.0568} & {\scriptsize{}-0.0048}\tabularnewline
{\scriptsize{}CaHf$_{0.50}$Ti$_{0.50}$Se$_{3}$} &  & {\scriptsize{}6.75} & {\scriptsize{}7.24} & {\scriptsize{}9.89} &  & {\scriptsize{}$P2_{1}/c$ (14)} & {\scriptsize{}483.76} & {\scriptsize{}-0.0911} & {\scriptsize{}-0.0038}\tabularnewline
\multicolumn{1}{c}{{\scriptsize{}CaHf$_{0.25}$Ti$_{0.75}$S$_{3}$}} &  & {\scriptsize{}6.42} & {\scriptsize{}6.86} & {\scriptsize{}9.36} &  & {\scriptsize{}$P\bar{1}$ (2)} & {\scriptsize{}411.86} & {\scriptsize{}-0.2184} & {\scriptsize{}-0.0051}\tabularnewline
{\scriptsize{}CaHf$_{0.25}$Ti$_{0.75}$Se$_{3}$} &  & {\scriptsize{}6.71} & {\scriptsize{}7.19} & {\scriptsize{}9.81} &  & {\scriptsize{}$P\bar{1}$ (2)} & {\scriptsize{}473.70} & {\scriptsize{}-0.0982} & {\scriptsize{}-0.0041}\tabularnewline
\multicolumn{1}{c}{{\scriptsize{}CaTiS$_{3}$}} &  & {\scriptsize{}6.38} & {\scriptsize{}6.79} & {\scriptsize{}9.27} &  & {\scriptsize{}$Pnma$ (62)} & {\scriptsize{}402.07} & {\scriptsize{}-0.0676} & {\scriptsize{}-}\tabularnewline
{\scriptsize{}CaTiSe$_{3}$} &  & {\scriptsize{}6.67} & {\scriptsize{}7.13} & {\scriptsize{}9.74} &  & {\scriptsize{}$Pnma$ (62)} & {\scriptsize{}463.27} & {\scriptsize{}-0.1007} & {\scriptsize{}-}\tabularnewline
\hline 
\end{tabular}{\scriptsize\par}
\end{table}

As the structural stability of perovskites plays a pivotal role in
determining their suitability for high-performance device applications,
we have evaluated the crystallographic stability of these materials.
The crystallographic stability of mixed and alloyed perovskites has
been analyzed by calculating Goldschmidt\textquoteright s tolerance
factor ($t$) \citep{chapter1-36,chapter1-37} and the octahedral
factor ($\mu$) \citep{chapter1-37,chapter1-38} (for details, see
the Supplemental Material). The calculated values of $t$ and $\mu$
are presented in Table S2, ranging from 0.873$-$0.920 and 0.306$-$0.386,
respectively, confirming the stability of the investigated perovskites.

\subsubsection{Thermodynamic stability:}

To evaluate the thermodynamic stability of the mixed and alloyed CPs,
their decomposition energies are calculated using the PBE xc functional.
In our study, the stability of CaHfS$_{3-x}$Se$_{x}$ and CaHf$_{1-y}$Ti$_{y}$X$_{3}$
compounds is estimated by calculating the energies needed to decompose
them into their binary phases, such as CaX, HfX$_{2}$, and TiX$_{2}$,
or into ternary phases like CaHfX$_{3}$ and CaTiX$_{3}$ (for details,
see the Supplemental Material). The positive decomposition energies
($E_{\mathrm{d}}$) indicate that these compounds will remain stable
and will not break down into their respective binary (B) or ternary
(T) phases. As shown in Table \ref{tab:1}, the compounds CaHfS$_{2}$Se
and CaHfSSe$_{2}$ do not decompose into ternary phases at 0 K, while
the remaining compounds have a high possibility of decomposing into
both phases at 0 K. Notably, the probability of decomposition into
binary phases is considerably higher than that into ternary phases
at 0 K. Nevertheless, there is still a chance for these compounds
to stabilize at elevated temperatures, as exemplified by the successful
synthesis of CaTiS$_{3}$ perovskite at 600-900°C in a vacuum \citep{chapter6-1},
suggesting their potential for higher temperature stability. In our
study, thermodynamic stability does not ensure the stability of these
materials; thus, we also assess their dynamical and mechanical stability.

\subsubsection{Dynamical stability:}

\begin{figure}[H]
\begin{centering}
\includegraphics[width=1\textwidth,height=1\textheight,keepaspectratio]{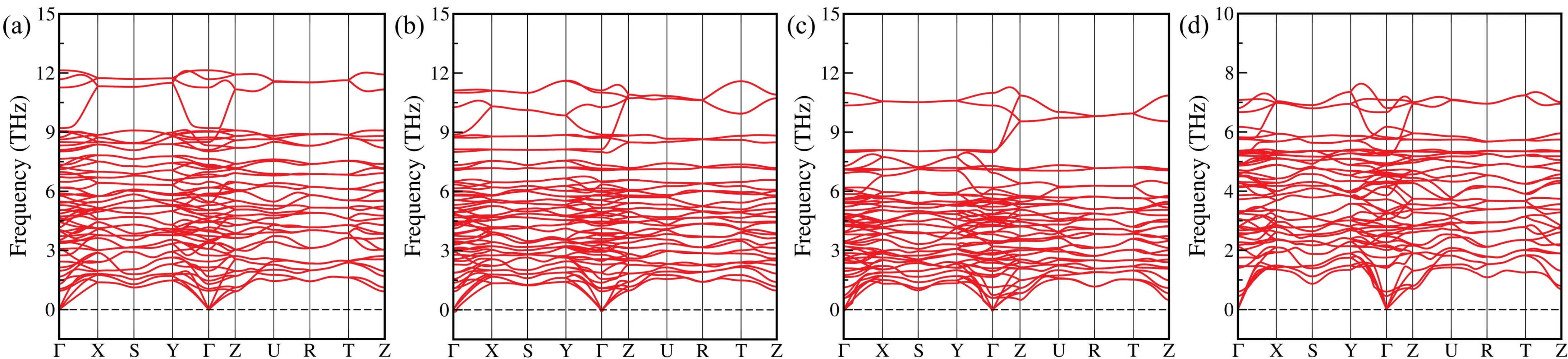}
\par\end{centering}
\caption{\label{fig:2}Phonon dispersion curves of (a) CaHfS$_{3}$, (b) CaHfS$_{2}$Se,
(c) CaHfSSe$_{2}$, and (d) CaHfSe$_{3}$ CPs, respectively, calculated
with the DFPT method.}
\end{figure}

The dynamical stability of the investigated CPs is also evaluated,
as it is a critical factor in determining material stability, directly
related to the behavior of phonon modes. To assess this, self-consistent
phonon calculations are performed using the DFPT method \citep{chapter1-60}.
The phonon dispersion curves of CaHfS$_{3-x}$Se$_{x}$ ($x$ = 0,
1, 2, 3), shown in Figure \ref{fig:2}(a)-(d), indicate that these
compounds are dynamically stable at 0 K. The remaining compounds are
found to be unstable at 0 K; however, they may achieve dynamical stability
at higher temperatures, as discussed before. For CaHfS$_{3-x}$Se$_{x}$
($x$ = 0, 1, 2, 3) perovskites, the structural symmetry results in
60 phonon modes corresponding to 20 atoms. Of these, 3 are acoustic,
while the remaining 57 are optical, categorized as low- and high-frequency
phonons, respectively. Notably, the highest optical frequency decreases
significantly with increasing Se ($x$) concentration.

\subsubsection{Mechanical stability and elastic properties:}

To assess the mechanical stability of these CPs, the second-order
elastic coefficients ($C_{ij}$) are calculated using the energy-strain
approach \citep{chapter1-47} (for details, see the Supplemental Material).
Since the CPs examined in this study belong to different crystal symmetry
classes, they exhibit varying numbers of independent elastic constants.
The calculated $C_{ij}$ values for these compounds are presented
in Table S4 of the Supplemental Material, and they satisfy the Born
stability criteria \citep{chapter1-47}. This certifies the mechanical
stability of these alloyed chalcogenide perovskites. The bulk modulus
($B$), shear modulus ($G$), Young's modulus ($Y$), and Poisson's
ratio ($\nu$) of all these materials are also determined using their
respective elastic coefficients \citep{chapter1-49,chapter1-50},
and are presented in Table S5 of the Supplemental Material. Our estimated
values for $B$ are significantly higher than those for $G$ in all
cases, indicating a stronger resistance to volumetric deformation
compared to shape deformation. The lower $G$ values reflect the flexibility
of the materials. The fragility of the materials is assessed using
Pugh's ratio ($B/G$) and Poisson's ratio ($\nu$) \citep{chapter1-51}.
The computed values of $B/G$ (> 1.75) and $\nu$ (> 0.26) indicate
that the investigated CPs exhibit ductile behavior, with the exceptions
of CaHfSSe$_{2}$ and CaTiSe$_{3}$, which are found to be brittle.
These particular mechanical properties make the selected CPs well-suited
for use in flexible and durable devices.

\subsection{Electronic properties:}

After ensuring the stability of these CPs, electronic structure calculations
are performed, as this is crucial for designing photoelectric devices.
Therefore, the electronic density of states (DOS), as well as band
structures and their nature, are evaluated to provide deep insights
into the electronic structure.

\begin{figure}[H]
\begin{centering}
\includegraphics[width=1\textwidth,height=1\textheight,keepaspectratio]{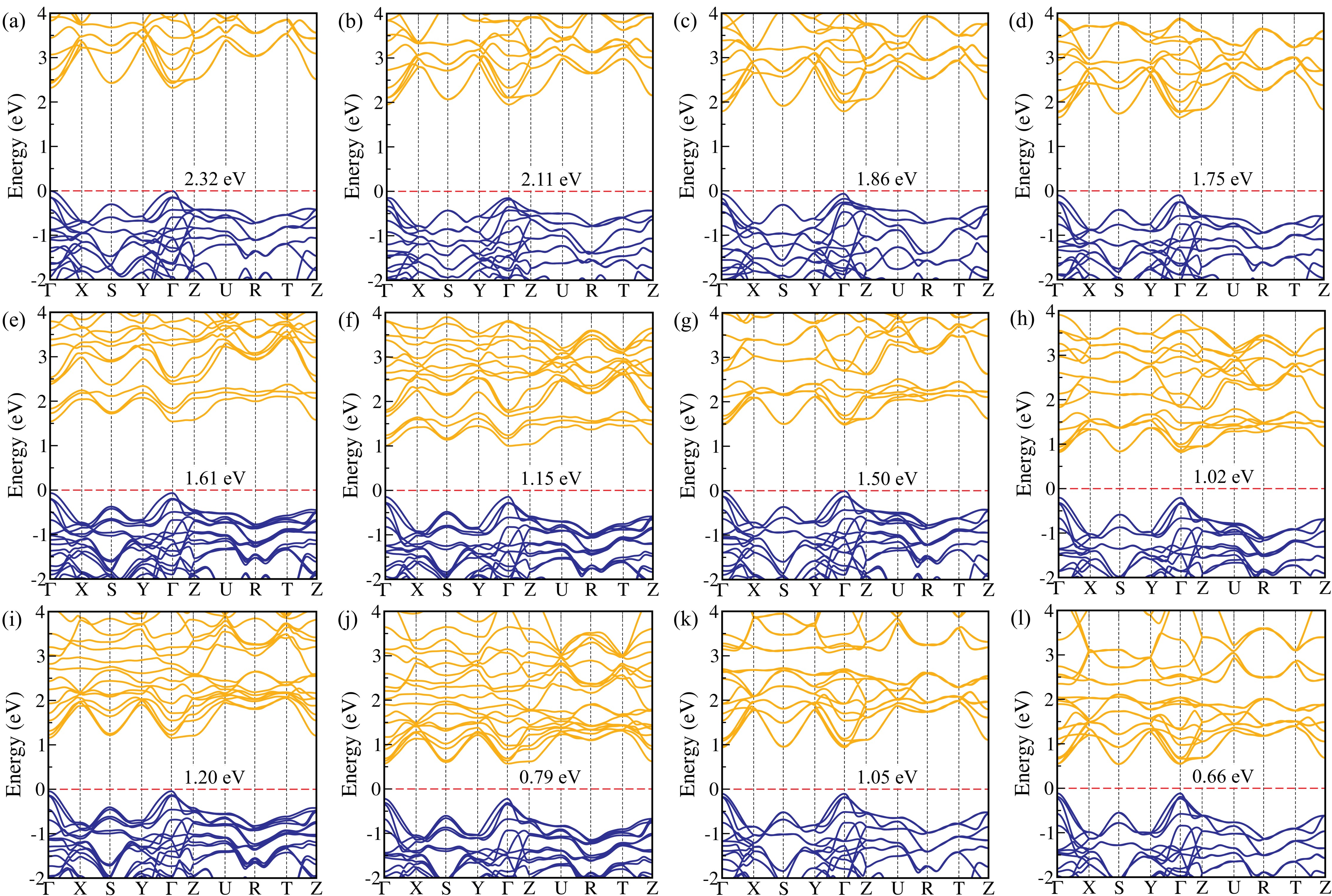}
\par\end{centering}
\caption{\label{fig:3}Electronic band structures of {[}(a)-(d){]} CaHfS$_{3-x}$Se$_{x}$
($x$ = 0, 1, 2, 3), {[}(e)-(f){]} CaHf$_{0.75}$Ti$_{0.25}$X$_{3}$
(X = S, Se), {[}(g)-(h){]} CaHf$_{0.50}$Ti$_{0.50}$X$_{3}$ (X =
S, Se), {[}(i)-(j){]} CaHf$_{0.25}$Ti$_{0.75}$X$_{3}$ (X = S, Se),
and {[}(k)-(l){]} CaTiX$_{3}$ (X = S, Se) CPs, respectively, obtained
using the HSE06 xc functional. The Fermi level is set to be zero and
marked by the dashed line.}

\end{figure}

Initially, electronic structure calculations for these alloyed CPs
are carried out using the widely adopted semilocal PBE xc functional,
both with and without including spin-orbit coupling (SOC) effects.
However, it is well-established that PBE underestimates the bandgap
of chalcogenide perovskites due to self-interaction error \citep{chapter1-63,chapter3-19,chapter5-16}.
Additionally, our results indicate that SOC has a minimal effect on
bandgaps (see Table S5 of the Supplemental Material). Therefore, the
hybrid HSE06 xc functional, along with the MBPT-based G$_{0}$W$_{0}$@PBE
method, has been utilized for precise bandgap estimation of these
CPs. The band structures of these compounds, as calculated using HSE06
and G$_{0}$W$_{0}$@PBE, are shown in Figure \ref{fig:3} and Figure
S2, respectively. It is observed that all compounds exhibit a direct
bandgap, as both the conduction band minimum (CBM) and valence band
maximum (VBM) are located at the same $\mathbf{k}$-point, specifically
at the $\Gamma$-point in the Brillouin zone. It is noteworthy that
CaTiSe$_{3}$ exhibits a slightly indirect bandgap, only 10 meV smaller
than the direct bandgap, attributed to the pronounced octahedral tilting
of the TiSe$_{6}$ units. The bandgaps of these CPs, estimated using
different functionals/methods, are presented in Table \ref{tab:3}
and show a close agreement with previous theoretical predictions and
available experimental data. For example, the G$_{0}$W$_{0}$@PBE
bandgap of CaTiS$_{3}$ is 1.83 eV, which closely matches its experimental
bandgap of 1.59 eV \citep{chapter6-1}. The difference in bandgap
values may be due to temperature, as the experimental measurement
was taken at 600°C in a vacuum. The direct bandgaps of these CPs,
calculated using HSE06 and G$_{0}$W$_{0}$@PBE, fall within the ranges
of 0.69$-$2.32 eV and 1.29$-$2.67 eV, respectively. The bandgaps
of these alloyed CPs indicate their suitability for photovoltaic applications,
aligning closely with those of conventional lead-based HPs (1.50$\lyxmathsym{\textendash}$3.13
eV) \citep{chapter5-11,chapter5-12,chapter5-13}.

{\footnotesize{}}
\begin{table}[H]
{\footnotesize{}\caption{\label{tab:3}Bandgap (in eV) of CPs calculated using the PBE, HSE06,
and G$_{0}$W$_{0}$@PBE method, respectively, as well as computed
effective mass of electron ($m_{e}^{*}$) and hole ($m_{h}^{*}$)
and reduced mass ($\mu^{*}$) average along the high-symmetry path
\textgreek{G}\textminus X, \textgreek{G}\textminus Y, and \textgreek{G}\textminus Z
directions. Here, $t$, $e$, $i$, and $d$ represent theoretical,
experimental, indirect, and direct bandgaps, respectively, and all
values of the effective mass are in terms of free-electron mass ($m_{0}$).}
}{\footnotesize\par}
\centering{}{\scriptsize{}}%
\begin{tabular}{cccccccc}
\hline 
{\scriptsize{}Configurations} & {\scriptsize{}PBE} & {\scriptsize{}HSE06} & {\scriptsize{}G$_{0}$W$_{0}$@PBE} & {\scriptsize{}Previous Work} & {\scriptsize{}$m_{e}^{*}$ ($m_{0}$)} & {\scriptsize{}$m_{h}^{*}$ ($m_{0}$)} & {\scriptsize{}$\mu^{*}$ ($m_{0}$)}\tabularnewline
\hline 
\multicolumn{1}{c}{{\scriptsize{}CaHfS$_{3}$}} & {\scriptsize{}1.52} & {\scriptsize{}2.32} & {\scriptsize{}2.67} & {\scriptsize{}2.46$_{t}$ \citep{chapter3-19}} & {\scriptsize{}0.431} & {\scriptsize{}0.380} & {\scriptsize{}0.202}\tabularnewline
{\scriptsize{}CaHfS$_{2}$Se} & {\scriptsize{}1.34} & {\scriptsize{}2.11} & {\scriptsize{}2.39} &  & {\scriptsize{}0.420} & {\scriptsize{}0.370} & {\scriptsize{}0.197}\tabularnewline
{\scriptsize{}CaHfSSe$_{2}$} & {\scriptsize{}1.15} & {\scriptsize{}1.86} & {\scriptsize{}2.09} &  & {\scriptsize{}0.361} & {\scriptsize{}0.309} & {\scriptsize{}0.166}\tabularnewline
{\scriptsize{}CaHfSe$_{3}$} & {\scriptsize{}1.05} & {\scriptsize{}1.75} & {\scriptsize{}1.92} &  & {\scriptsize{}0.358} & {\scriptsize{}0.274} & {\scriptsize{}0.155}\tabularnewline
\multicolumn{1}{c}{{\scriptsize{}CaHf$_{0.75}$Ti$_{0.25}$S$_{3}$}} & {\scriptsize{}0.80} & {\scriptsize{}1.61} & {\scriptsize{}2.48} &  & {\scriptsize{}0.552} & {\scriptsize{}0.347} & {\scriptsize{}0.213}\tabularnewline
{\scriptsize{}CaHf$_{0.75}$Ti$_{0.25}$Se$_{3}$} & {\scriptsize{}0.43} & {\scriptsize{}1.15} & {\scriptsize{}1.89} &  & {\scriptsize{}0.444} & {\scriptsize{}0.243} & {\scriptsize{}0.157}\tabularnewline
\multicolumn{1}{c}{{\scriptsize{}CaHf$_{0.50}$Ti$_{0.50}$S$_{3}$}} & {\scriptsize{}0.72} & {\scriptsize{}1.50} & {\scriptsize{}2.34} &  & {\scriptsize{}0.463} & {\scriptsize{}0.312} & {\scriptsize{}0.186}\tabularnewline
{\scriptsize{}CaHf$_{0.50}$Ti$_{0.50}$Se$_{3}$} & {\scriptsize{}0.37} & {\scriptsize{}1.02} & {\scriptsize{}1.64} &  & {\scriptsize{}0.345} & {\scriptsize{}0.202} & {\scriptsize{}0.127}\tabularnewline
\multicolumn{1}{c}{{\scriptsize{}CaHf$_{0.25}$Ti$_{0.75}$S$_{3}$}} & {\scriptsize{}0.46} & {\scriptsize{}1.20} & {\scriptsize{}1.99} &  & {\scriptsize{}0.406} & {\scriptsize{}0.273} & {\scriptsize{}0.163}\tabularnewline
{\scriptsize{}CaHf$_{0.25}$Ti$_{0.75}$Se$_{3}$} & {\scriptsize{}0.16} & {\scriptsize{}0.79} & {\scriptsize{}1.49} &  & {\scriptsize{}0.277} & {\scriptsize{}0.160} & {\scriptsize{}0.101}\tabularnewline
\multicolumn{1}{c}{{\scriptsize{}CaTiS$_{3}$}} & {\scriptsize{}0.35} & {\scriptsize{}1.05} & {\scriptsize{}1.83} & {\scriptsize{}1.59$_{e}$ \citep{chapter6-1}} & {\scriptsize{}0.264} & {\scriptsize{}0.224} & {\scriptsize{}0.121}\tabularnewline
{\scriptsize{}CaTiSe$_{3}$} & {\scriptsize{}0.07$^{i}$ (0.08$^{d}$)} & {\scriptsize{}0.66$^{i}$ (0.69$^{d}$)} & {\scriptsize{}1.17$^{i}$ (1.29$^{d}$)} &  & {\scriptsize{}0.162} & {\scriptsize{}0.103} & {\scriptsize{}0.063}\tabularnewline
\hline 
\end{tabular}{\scriptsize\par}
\end{table}
{\footnotesize\par}

\begin{figure}[H]
\begin{centering}
\includegraphics[width=1\textwidth,height=1\textheight,keepaspectratio]{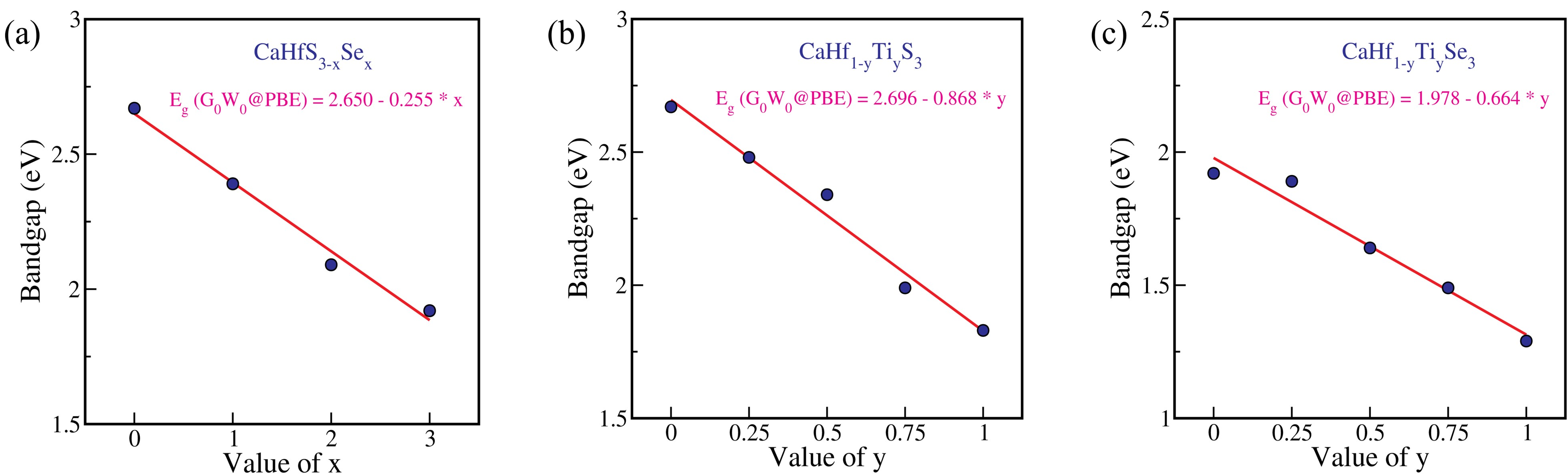}
\par\end{centering}
\caption{\label{fig4}Variation of the calculated G$_{0}$W$_{0}$@PBE bandgap
with the value of $x$ and $y$ in (a) CaHfS$_{3-x}$Se$_{x}$, (b)
CaHf$_{1-y}$Ti$_{y}$S$_{3}$, and (c) CaHf$_{1-y}$Ti$_{y}$Se$_{3}$
CPs.}

\end{figure}

Furthermore, the variation in bandgap for the mixed CaHfS$_{3-x}$Se$_{x}$
($x$ = 0, 1, 2, 3) and alloyed CaHf$_{1-y}$Ti$_{y}$X$_{3}$ ($y$
= 0, 0.25, 0.5, 0.75, 1; X = S, Se) CPs as a function of $x$ and
$y$ is shown in Figure \ref{fig4} and fitted for G$_{0}$W$_{0}$@PBE
method. It is observed that the bandgap of these compounds decreases
linearly as the values of $x$ (Se) and $y$ (Ti) increase, with deviations
of 255 meV, 868 meV, and 664 meV for CaHfS$_{3-x}$Se$_{x}$, CaHf$_{1-y}$Ti$_{y}$S$_{3}$,
and CaHf$_{1-y}$Ti$_{y}$Se$_{3}$, respectively. These trends may
aid in predicting the bandgap of mixed and alloyed compounds with
different $x$ and $y$ values, which is often encountered during
the synthesis of these materials \citep{chapter6-2,chapter6-3}.

To provide a more comprehensive analysis of the electronic band structures
of these alloyed compounds, both the total (TDOS) and partial (PDOS)
density of states have been calculated using the HSE06 xc functional.
The results are illustrated in Figure S1 of the Supplemental Material.
For CaHfS$_{3}$, the VBM primarily consists of the S-3$p$ orbitals
with a small contribution from Hf-5$d$ orbitals, whereas the CBM
is predominantly derived from the Hf-5$d$ orbitals with a minor contribution
from S-3$p$ orbitals. Similarly, as the Se content increases in CaHfS$_{3-x}$Se$_{x}$,
the contribution of Se-4$p$ orbitals becomes more significant compared
to S-3$p$ orbitals. Since the S-3$p$ orbitals are at a lower energy
level than the Se-4$p$ orbitals, the higher energy of the Se-4$p$
orbitals reduces the energy difference between the VBM (primarily
Se-$4p$) and the CBM. This leads to a decrease in bandgap as the
Se content increases in CaHfS$_{3-x}$Se$_{x}$ compounds. Conversely,
when the Ti content in CaHf$_{1-y}$Ti$_{y}$X$_{3}$ (X = S, Se)
alloys increases, the contribution of Ti-3$d$ orbitals increases
with respect to Hf-5$d$ orbitals. The Ti-3$d$ orbitals are more
localized than the Hf-5$d$ orbitals, so alloying with Ti lowers the
energy of the CBM and reduces the bandgaps of these alloys.

Further, to gain a deep insight into charge carrier transport, we
computed the effective masses of electrons ($m_{e}^{*}$) and holes
($m_{h}^{*}$) for all the investigated compounds by fitting the $E-k$
dispersion curves obtained from G$_{0}$W$_{0}$@PBE band structures.
The effective masses of these CPs are determined using the relation,
$\text{\ensuremath{m^{*}=\hbar^{2}\left[\partial^{2}E(k)/\partial k^{2}\right]^{-1}}}$
and the corresponding values are presented in Table \ref{tab:3}.
Note that, effective masses of the charge carriers are estimated along
three different directions: \textgreek{G}$-$X, \textgreek{G}$-$Y,
and \textgreek{G}$-$Z, and then calculate their harmonic mean values
using Eq. \ref{eq:2} (for details, see the Supplemental Material).
From Table \ref{tab:3}, it is evident that $m_{e}^{*}$ and $m_{h}^{*}$
lie in the range of 0.162$-$0.552 and 0.103$-$0.380, respectively.
These values suggest high ambipolar carrier mobility, which implies
enhanced charge carrier transport in these materials.

\subsection{Optical Properties:}

To gain a comprehensive understanding of the suitability of a material
for optoelectronic applications, an in-depth analysis of its optical
properties, including the dielectric function and absorption coefficient,
is indispensable. To enhance the reliability of our predictions, we
have employed MBPT-based GW-BSE simulations, which explicitly incorporate
electron-hole interactions. GW calculations determine the fundamental
bandgap, which is considered more accurate and closely aligned with
results from experimental photoelectron spectroscopy (PES) and inverse
photoelectron spectroscopy (IPES) \citep{chapter1-69,chapter1-70}.
On the other hand, BSE calculations predict the optical bandgap, providing
results comparable to experimental optical absorption spectroscopy
\citep{chapter1-67,chapter1-68}. To obtain the optical response of
these CPs, a single-shot GW (G$_{0}$W$_{0}$) calculation is first
performed based on the PBE functional. Following this, the Bethe-Salpeter
equation (BSE) is solved using the G$_{0}$W$_{0}$@PBE results. The
optical response is assessed by calculating the frequency-dependent
dielectric function, $\varepsilon(\omega)$, expressed as $\varepsilon(\omega)$
= {[}Re($\varepsilon$){]} + i{[}Im($\varepsilon$){]}, where {[}Re($\varepsilon$){]}
denotes the real part, and {[}Im($\varepsilon$){]} represents the
imaginary part of the dielectric function. 

\begin{figure}[H]
\begin{centering}
\includegraphics[width=1\textwidth,height=1\textheight,keepaspectratio]{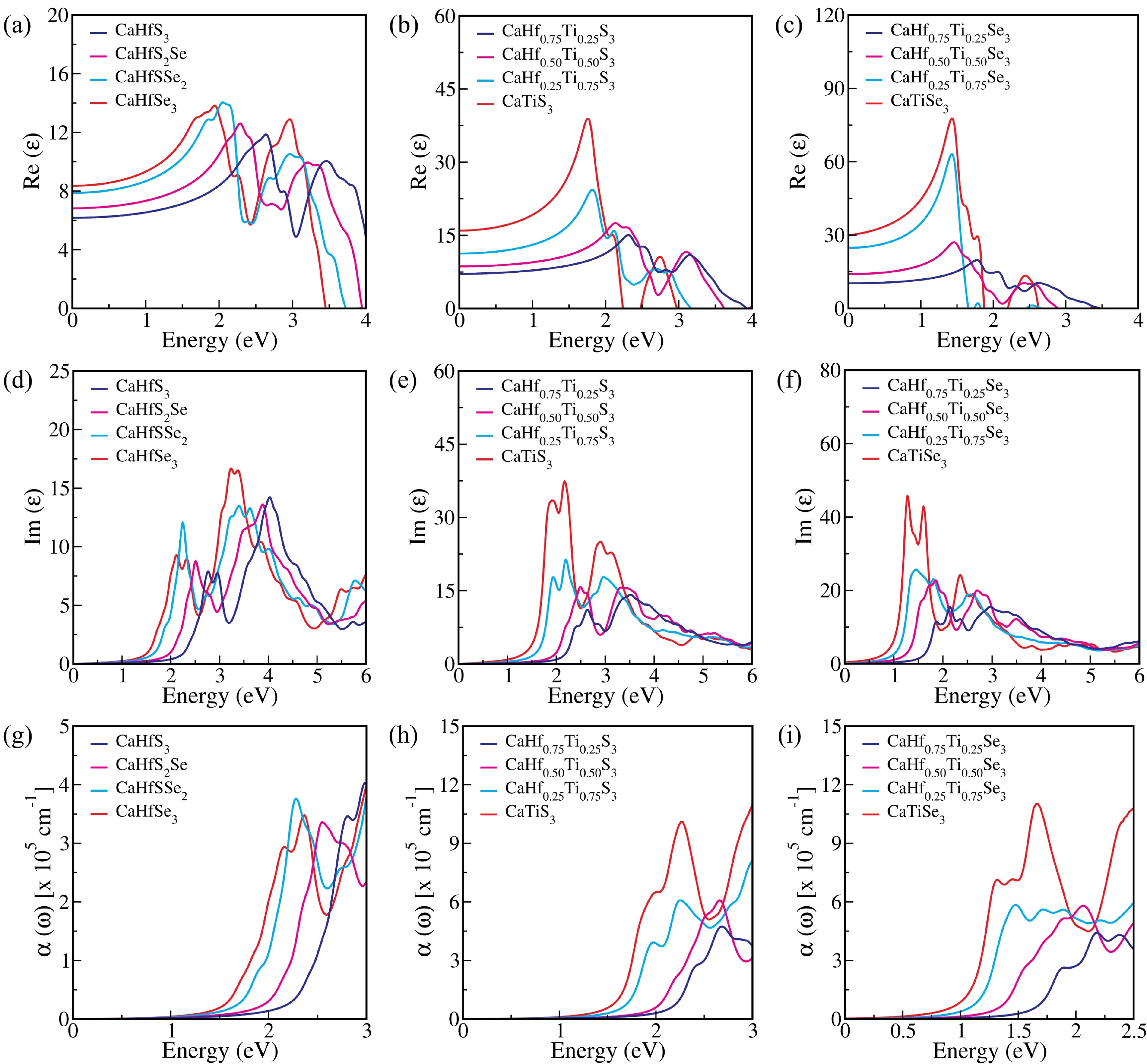}
\par\end{centering}
\caption{\label{fig5}Real {[}Re($\varepsilon$){]} and imaginary {[}Im($\varepsilon$){]}
part of the electronic dielectric function, and absorption coefficient
{[}$\alpha(\omega)${]} of CaHfS$_{3-x}$Se$_{x}$ ($x$ = 0, 1, 2,
3) and CaHf$_{1-y}$Ti$_{y}$X$_{3}$ ($y$ = 0, 0.25, 0.5, 0.75,
1; X = S, Se) CPs, obtained using the BSE@G$_{0}$W$_{0}$@PBE method.}

\end{figure}

The real part of the dielectric function, {[}Re($\varepsilon$){]},
represents how a material responds to an electric field through polarization.
Higher values of {[}Re($\varepsilon$){]} indicate stronger polarization,
which influences the material's optical properties, including its
refractive index and light absorption. The calculated {[}Re($\varepsilon$){]}
of the investigated CPs obtained using BSE@G$_{0}$W$_{0}$@PBE is
depicted in Figure \ref{fig5}(a)-(c). Furthermore, the real part
of the dielectric function {[}Re($\varepsilon$){]}, at zero energy,
is commonly referred to as the electronic or optical dielectric constant
($\varepsilon_{\infty}$)\textemdash characterizes the dielectric
screening effect during electron-hole Coulomb interactions. Our results
show that the value of $\varepsilon_{\infty}$ increases with the
incorporation of Ti and Se in these CPs. This trend indicates reduced
charge carrier recombination rates and improved optoelectronic performance
in these materials, as a higher $\varepsilon_{\infty}$ suggests stronger
electronic screening and enhanced dielectric response. Notably, $\varepsilon_{\infty}$
for CaTiSe$_{3}$ reaches 29.93, the highest among the studied compounds,
highlighting its superior potential for optoelectronic applications.

On the other hand, the imaginary part of the dielectric function,
{[}Im($\varepsilon$){]}, plays a key role in determining the linear
absorption properties of a material. Figure \ref{fig5}(d)-(f) represents
the {[}Im($\varepsilon$){]} of these CPs calculated using BSE@G$_{0}$W$_{0}$@PBE.
It has been observed that both the absorption onset and the position
of the first peak ($E_{o}$) of these CPs exhibit a gradual red shift
with increasing concentrations of Ti and Se. This shift corresponds
to a decrease in the quasiparticle (QP) bandgap, as detailed in Table
\ref{tab:3}. For example, the $E_{o}$ value for CaHfS$_{3}$ is
2.45 eV, whereas it decreases to 1.27 eV for CaTiSe$_{3}$. It should
be emphasized that these CPs exhibit an absorption onset from the
near-infrared to the visible region, making them advantageous for
solar cell applications.

One of the most crucial parameters for evaluating a material's suitability
for photovoltaic applications is the absorption coefficient. This
metric is essential for determining the material's potential for optimal
solar energy conversion efficiency. Consequently, the absorption coefficient
{[}$\alpha(\omega)${]} for each of the examined CPs is calculated
using the following formula \citep{chapter1-60},

\begin{equation}
\alpha(\omega)=\sqrt{2}\omega\text{\ensuremath{\left[\sqrt{\{\mathrm{Re}(\varepsilon)\}^{2}+\{\mathrm{Im}(\varepsilon)\}^{2}}-\mathrm{Re}(\varepsilon)\right]^{1/2}.}}
\end{equation}

Generally, a standard $\alpha(\omega)$ in the order of 10$^{5}$
cm$^{-1}$ is preferred for direct semiconductors \citep{chapter5-16}.
Based on our BSE@G$_{0}$W$_{0}$@PBE results, all of the studied
perovskites meet this criterion {[}see Figure \ref{fig5}(g)-(i){]},
suggesting their potential for use in photovoltaic devices.

\subsection{Excitonic Properties:}

In addition to the optical properties, various excitonic parameters,
including exciton binding energy ($E_{B}$), exciton radius ($r_{exc}$),
and exciton lifetime ($\tau_{exc}$), are also calculated for these
perovskites. Excitons are bound states formed by an electron ($e$)
and a hole ($h$) upon light absorption in a material, and the characteristics
of these excitons significantly influence the photovoltaic performance
of this material. The energy needed to separate an exciton into a
free $e$$-$$h$ pair is called the exciton binding energy ($E_{B}$).
In photovoltaics, a lower $E_{B}$ promotes easier separation of $e$$-$$h$
pairs, which enhances photoelectric conversion efficiency. Theoretically,
$E_{B}$ is estimated from first-principles BSE calculations as the
difference between the QP bandgap, $E_{g}^{d}$ (direct G$_{0}$W$_{0}$@PBE
bandgap), and the optical bandgap, $E_{o}$ (first peak position from
BSE@G$_{0}$W$_{0}$@PBE) \citep{chapter2-38,chapter5-18}. The $E_{B}$
values, calculated using the first-principles BSE method for the systems
under investigation, are shown in Table \ref{tab:4}, ranging from
0.016 to 0.221 eV. This indicates that these materials can achieve
a comparable or even broader range of $E_{B}$ values than conventional
lead-based HPs (0.01$-$0.10 eV) \citep{chapter5-11,chapter5-12,chapter5-14}.
The higher $E_{B}$ values observed in some of our systems suggest
substantial excitonic effects, which can be advantageous for applications
requiring enhanced exciton binding, such as excitonic solar cells
and light-emitting devices. Additionally, the ability to access both
low and high $E_{B}$ regimes offers tunability, making these materials
versatile candidates for diverse optoelectronic applications. The
$E_{B}$ values of these CPs are also found to decrease with increasing
Ti concentration, which can be attributed to higher electronic dielectric
constants ($\varepsilon_{\infty}$) and lower effective masses (see
Equation \ref{eq:1}).

In our study, Tables \ref{tab:4} and \ref{tab:5} suggest that $E_{B}\gg\hbar\omega_{LO}$
for most of the systems, where $\omega_{LO}$ is the characteristic
longitudinal optical phonon frequency. In this case, the electronic
contribution to the dielectric screening dominates over the ionic
contribution, allowing the ionic part to be neglected, which leads
to no change in $E_{B}$ \citep{chapter1-65,chapter1-66}. Further,
the hydrogenic Wannier-Mott (WM) model \citep{chapter1-63,chapter5-16}
is employed to determine the $E_{B}$ for these mixed and alloyed
perovskites using Eq. \ref{eq:1}. In this model, $\mathrm{\varepsilon_{eff}}$
is positioned between the electronic ($\varepsilon_{\infty}$) and
static ($\varepsilon_{static}=\varepsilon_{\infty}+\varepsilon_{ion}$)
dielectric constant, where $\varepsilon_{ion}$ represents the ionic
contribution to the dielectric function (for details, see the Supplemental
Material). The values of $\varepsilon_{\infty}$ and $\varepsilon_{ion}$
are derived from the BSE and DFPT methods, respectively. The electronic
and static dielectric constants establish the upper ($E_{Bu}$) and
lower ($E_{Bl}$) bounds of exciton binding energy (for details, see
the Supplemental Material). Our study also reveals that the upper
bound values are smaller but align more closely with the $E_{B}$
calculated using the standard BSE method than the lower bound values.
Nevertheless, the overall trend remains consistent, indicating that
the electronic contribution to dielectric screening is more prominent
than the ionic contribution, and $\mathrm{\varepsilon_{eff}}$ $\rightarrow$
$\varepsilon_{\infty}$ in these CPs.

\begin{table}[H]
\caption{\label{tab:4}Calculated excitonic parameters, phonon screening corrections
($\Delta E_{B}^{ph}$), and modified exciton binding energy ($E_{B}^{\prime}$)
of chalcogenide perovskites.}

\centering{}{\footnotesize{}}%
\begin{tabular}{ccccccc}
\hline 
{\footnotesize{}Configurations} & {\footnotesize{}$E_{B}$ (eV)} & {\footnotesize{}$T_{exc}$ (K)} & {\footnotesize{}$r_{exc}$ (nm)} & {\footnotesize{}$|\phi_{n}(0)|^{2}$ (10$^{26}$ m$^{-3}$)} & {\footnotesize{}$\Delta E_{B}^{ph}$ (meV)} & {\footnotesize{}$E_{B}^{\prime}$ (eV)}\tabularnewline
\hline 
{\footnotesize{}CaHfS$_{3}$} & {\footnotesize{}0.221} & {\footnotesize{}2562} & {\footnotesize{}1.62} & {\footnotesize{}0.75} & {\footnotesize{}-16.60} & {\footnotesize{}0.204}\tabularnewline
{\footnotesize{}CaHfS$_{2}$Se} & {\footnotesize{}0.183} & {\footnotesize{}2122} & {\footnotesize{}1.83} & {\footnotesize{}0.52} & {\footnotesize{}-15.42} & {\footnotesize{}0.168}\tabularnewline
{\footnotesize{}CaHfSSe$_{2}$} & {\footnotesize{}0.157} & {\footnotesize{}1820} & {\footnotesize{}2.51} & {\footnotesize{}0.20} & {\footnotesize{}-12.91} & {\footnotesize{}0.144}\tabularnewline
{\footnotesize{}CaHfSe$_{3}$} & {\footnotesize{}0.046} & {\footnotesize{}533} & {\footnotesize{}2.85} & {\footnotesize{}0.14} & {\footnotesize{}-9.90} & {\footnotesize{}0.036}\tabularnewline
{\footnotesize{}CaHf$_{0.75}$Ti$_{0.25}$S$_{3}$} & {\footnotesize{}0.133} & {\footnotesize{}1542} & {\footnotesize{}1.76} & {\footnotesize{}0.58} & {\footnotesize{}-15.70} & {\footnotesize{}0.117}\tabularnewline
{\footnotesize{}CaHf$_{0.75}$Ti$_{0.25}$Se$_{3}$} & {\footnotesize{}0.063} & {\footnotesize{}730} & {\footnotesize{}3.46} & {\footnotesize{}0.08} & {\footnotesize{}-10.89} & {\footnotesize{}0.052}\tabularnewline
{\footnotesize{}CaHf$_{0.50}$Ti$_{0.50}$S$_{3}$} & {\footnotesize{}0.116} & {\footnotesize{}1345} & {\footnotesize{}2.46} & {\footnotesize{}0.21} & {\footnotesize{}-15.17} & {\footnotesize{}0.101}\tabularnewline
{\footnotesize{}CaHf$_{0.50}$Ti$_{0.50}$Se$_{3}$} & {\footnotesize{}0.059} & {\footnotesize{}684} & {\footnotesize{}5.84} & {\footnotesize{}0.02} & {\footnotesize{}-10.27} & {\footnotesize{}0.049}\tabularnewline
{\footnotesize{}CaHf$_{0.25}$Ti$_{0.75}$S$_{3}$} & {\footnotesize{}0.080} & {\footnotesize{}928} & {\footnotesize{}3.66} & {\footnotesize{}0.06} & {\footnotesize{}-14.85} & {\footnotesize{}0.065}\tabularnewline
{\footnotesize{}CaHf$_{0.25}$Ti$_{0.75}$Se$_{3}$} & {\footnotesize{}0.051} & {\footnotesize{}591} & {\footnotesize{}12.97} & {\footnotesize{}0.001} & {\footnotesize{}-9.59} & {\footnotesize{}0.041}\tabularnewline
{\footnotesize{}CaTiS$_{3}$} & {\footnotesize{}0.022} & {\footnotesize{}255} & {\footnotesize{}6.98} & {\footnotesize{}0.009} & {\footnotesize{}-11.48} & {\footnotesize{}0.011}\tabularnewline
{\footnotesize{}CaTiSe$_{3}$} & {\footnotesize{}0.016} & {\footnotesize{}186} & {\footnotesize{}25.13} & {\footnotesize{}0.0002} & {\footnotesize{}-7.85} & {\footnotesize{}0.008}\tabularnewline
\hline 
\end{tabular}{\footnotesize\par}
\end{table}

Using the aforementioned quantities ($E_{B}$, $\varepsilon_{\infty}$,
and $\mu^{*}$), several excitonic parameters are also computed, including
excitonic temperature ($T_{exc}$), radius ($r_{exc}$), and probability
of wave function ($|\phi_{n}(0)|^{2}$) for $e-h$ pair at zero separation.
$T_{exc}$ represents the maximum temperature at which an exciton
remains stable, and the thermal energy needed to separate an exciton
is given by $E_{B}=k_{B}T_{exc}$, where $k_{B}$ is the Boltzmann
constant. Furthermore, $r_{exc}$ and $|\phi_{n}(0)|^{2}$ are calculated
using Eqs. \ref{eq:5} and \ref{eq:6}, respectively. The values of
these parameters are listed in Table \ref{tab:4}, and the inverse
of $|\phi_{n}(0)|^{2}$ can be used to qualitatively characterize
the exciton lifetime ($\tau_{exc}$; for details, see the Supplemental
Material). Table \ref{tab:4} indicates that the $\tau_{exc}$ values
for the investigated CPs increase with the rising concentration of
Ti in conjunction with Se. A longer exciton lifetime is associated
with a lower carrier recombination rate, which improves the quantum
yield and conversion efficiency.

Our results show that the upper bound values of $E_{B}$, estimated
using the WM model, are smaller but more closely aligned with the
$E_{B}$ values calculated using the standard BSE method. One reason
is that the standard BSE (ab initio) approach considers only electronic
screening when constructing the $e-h$ kernel for calculating the
$E_{B}$. However, static screening does not account for electron-phonon
coupling, which can be crucial in certain materials, particularly
those with significant electron-phonon interactions or where phonons
are key to determining optoelectronic properties. In a recent study,
Filip and colleagues \citep{chapter3-39} incorporated phonon screening
into $E_{B}$ by considering four distinct material parameters: $\mu^{*}$,
$\varepsilon_{\infty}$, $\varepsilon_{static}$, and $\omega_{LO}$.
The phonon-screening correction ($\Delta E_{B}^{ph}$), assuming isotropic
and parabolic electronic band dispersion, is computed through Eq.
\ref{eq:4}, with the corresponding values for these compounds provided
in Table \ref{tab:4}. The contribution of phonon screening is found
to reduce $E_{B}$ by 7.85 meV to 16.60 meV, which is significant
for certain materials and, therefore, cannot be overlooked. After
incorporating the phonon-screening correction, the modified exciton
binding energy ($E_{B}^{\prime}$) ranges from 0.008 to 0.204 eV,
indicating that these perovskite alloys are promising materials for
photovoltaic applications.

\subsection{Polaronic Properties:}

To further enhance our understanding of the fundamental limits of
carrier mobility for these CPs, it would be advantageous to make predictions
based on first-principles calculations \citep{chapter2-20,chapter2-21}.
In polar semiconductors, the scattering mechanism near room temperature
is dominated by the interaction between charge carriers and the macroscopic
electric field generated by longitudinal optical (LO) phonons \citep{chapter5-16,chapter5-18}.
Therefore, when calculating the mobility using theoretical approaches,
it is crucial to consider the polaron state, which arises from the
strong interactions between charge carriers and phonons, rather than
focusing solely on the free carrier state. Fr\"ohlich formulated a Hamiltonian
to describe the interaction between independent charge carriers (i.e.,
those with low density) and polar optical phonons \citep{chapter2-51}.
This interaction is characterized by the dimensionless Fr\"ohlich parameter,
$\alpha$, as defined in Eq. \ref{eq:7}. The computed carrier-phonon
coupling constants ($\alpha$) are listed in Table \ref{tab:5}. Strong
carrier-phonon coupling is typically indicated by $\alpha>10$, whereas
$\alpha\ll1$ usually suggests weak coupling \citep{chapter2-20}.
Our results indicate that the polaron in these materials resides in
the weak to intermediate coupling regime ($\alpha$ = 0.34$-$2.68).
In our study, CaHf$_{0.25}$Ti$_{0.75}$Se$_{3}$ and CaTiSe$_{3}$
exhibit smaller electron-phonon coupling relative to the other materials.
The reduced carrier-phonon coupling can be ascribed to factors such
as a lower effective mass of the carriers and a higher electronic
dielectric constant. Additionally, the Debye temperature ($\theta_{D}$)
for these CPs is calculated and found to be lower than room temperature
(see Table \ref{tab:5}), indicating a significant interaction between
carriers and phonons.

Polaron formation refers to the interaction of a charge carrier (electron
or hole) with its surrounding lattice, causing a local distortion.
This interaction can reduce the QP energies, meaning that both electrons
and holes lose energy when forming polarons. The polaron energy ($E_{p}$)
can be determined from $\alpha$ using Eq. \ref{eq:8}. The QP gap,
arising from the polaron energy of electrons and holes (see Table
\ref{tab:5}), is also computed and compared with the $E_{B}$ from
Table \ref{tab:4}. From this comparison, we observe that, for the
investigated CPs, the energy of charge-separated polaronic states
is lower than that of bound exciton states, except in the cases of
CaHfSe$_{3}$ and CaTiS$_{3}$. For example, the QP gap of CaHf$_{0.25}$Ti$_{0.75}$Se$_{3}$
is 12.49 meV, while the $E_{B}$ value for this compound is 51 meV.
This suggests that in most systems, charge-separated polaronic states
(where electrons and holes are further apart) are less stable than
bound excitons (where they remain closely associated).

\begin{table}[H]
\caption{\label{tab:5}Polaron parameters for electrons ($e$) and holes ($h$)
in chalcogenide perovskites.}

\centering{}{\footnotesize{}}%
\begin{tabular}{cccccccccccccc}
\hline 
\multirow{2}{*}{{\footnotesize{}Configurations}} & \multirow{2}{*}{{\footnotesize{}$\omega_{LO}$ (THz)}} & \multirow{2}{*}{{\footnotesize{}$\theta_{D}$ (K)}} & \multicolumn{2}{c}{{\footnotesize{}$\alpha$}} & \multirow{2}{*}{} & \multicolumn{2}{c}{{\footnotesize{}$E_{p}$ (meV)}} & \multirow{2}{*}{} & \multicolumn{2}{c}{{\footnotesize{}$m_{p}/m^{*}$}} & \multirow{2}{*}{} & \multicolumn{2}{c}{{\footnotesize{}$\mu_{p}$ (cm$^{2}$V$^{-1}$s$^{-1}$)}}\tabularnewline
\cline{4-5} \cline{5-5} \cline{7-8} \cline{8-8} \cline{10-11} \cline{11-11} \cline{13-14} \cline{14-14} 
 &  &  & {\footnotesize{}$e$} & {\footnotesize{}$h$} &  & {\footnotesize{}$e$} & {\footnotesize{}$h$} &  & {\footnotesize{}$e$} & {\footnotesize{}$h$} &  & {\footnotesize{}$e$} & {\footnotesize{}$h$}\tabularnewline
\hline 
{\footnotesize{}CaHfS$_{3}$} & {\footnotesize{}5.14} & {\footnotesize{}247} & {\footnotesize{}2.22} & {\footnotesize{}2.08} &  & {\footnotesize{}48.55} & {\footnotesize{}45.41} &  & {\footnotesize{}1.49} & {\footnotesize{}1.46} &  & {\footnotesize{}27.83} & {\footnotesize{}34.59}\tabularnewline
{\footnotesize{}CaHfS$_{2}$Se} & {\footnotesize{}4.91} & {\footnotesize{}236} & {\footnotesize{}1.99} & {\footnotesize{}1.87} &  & {\footnotesize{}41.45} & {\footnotesize{}38.90} &  & {\footnotesize{}1.43} & {\footnotesize{}1.40} &  & {\footnotesize{}33.84} & {\footnotesize{}41.78}\tabularnewline
{\footnotesize{}CaHfSSe$_{2}$} & {\footnotesize{}4.15} & {\footnotesize{}199} & {\footnotesize{}1.72} & {\footnotesize{}1.59} &  & {\footnotesize{}30.19} & {\footnotesize{}27.86} &  & {\footnotesize{}1.36} & {\footnotesize{}1.33} &  & {\footnotesize{}51.02} & {\footnotesize{}65.94}\tabularnewline
{\footnotesize{}CaHfSe$_{3}$} & {\footnotesize{}3.55} & {\footnotesize{}170} & {\footnotesize{}1.76} & {\footnotesize{}1.54} &  & {\footnotesize{}26.43} & {\footnotesize{}23.07} &  & {\footnotesize{}1.37} & {\footnotesize{}1.31} &  & {\footnotesize{}53.60} & {\footnotesize{}82.70}\tabularnewline
{\footnotesize{}CaHf$_{0.75}$Ti$_{0.25}$S$_{3}$} & {\footnotesize{}4.39} & {\footnotesize{}211} & {\footnotesize{}2.68} & {\footnotesize{}2.13} &  & {\footnotesize{}50.33} & {\footnotesize{}39.74} &  & {\footnotesize{}1.63} & {\footnotesize{}1.47} &  & {\footnotesize{}17.70} & {\footnotesize{}39.09}\tabularnewline
{\footnotesize{}CaHf$_{0.75}$Ti$_{0.25}$Se$_{3}$} & {\footnotesize{}3.22} & {\footnotesize{}155} & {\footnotesize{}1.91} & {\footnotesize{}1.42} &  & {\footnotesize{}26.07} & {\footnotesize{}19.27} &  & {\footnotesize{}1.41} & {\footnotesize{}1.29} &  & {\footnotesize{}40.80} & {\footnotesize{}107.60}\tabularnewline
{\footnotesize{}CaHf$_{0.50}$Ti$_{0.50}$S$_{3}$} & {\footnotesize{}4.24} & {\footnotesize{}204} & {\footnotesize{}2.07} & {\footnotesize{}1.70} &  & {\footnotesize{}37.27} & {\footnotesize{}30.47} &  & {\footnotesize{}1.45} & {\footnotesize{}1.36} &  & {\footnotesize{}30.93} & {\footnotesize{}59.47}\tabularnewline
{\footnotesize{}CaHf$_{0.50}$Ti$_{0.50}$Se$_{3}$} & {\footnotesize{}2.90} & {\footnotesize{}139} & {\footnotesize{}1.36} & {\footnotesize{}1.04} &  & {\footnotesize{}16.61} & {\footnotesize{}12.65} &  & {\footnotesize{}1.27} & {\footnotesize{}1.20} &  & {\footnotesize{}83.76} & {\footnotesize{}308.68}\tabularnewline
{\footnotesize{}CaHf$_{0.25}$Ti$_{0.75}$S$_{3}$} & {\footnotesize{}5.14} & {\footnotesize{}247} & {\footnotesize{}1.16} & {\footnotesize{}0.95} &  & {\footnotesize{}25.04} & {\footnotesize{}20.46} &  & {\footnotesize{}1.23} & {\footnotesize{}1.18} &  & {\footnotesize{}68.39} & {\footnotesize{}123.99}\tabularnewline
{\footnotesize{}CaHf$_{0.25}$Ti$_{0.75}$Se$_{3}$} & {\footnotesize{}3.80} & {\footnotesize{}182} & {\footnotesize{}0.45} & {\footnotesize{}0.34} &  & {\footnotesize{}7.12} & {\footnotesize{}5.37} &  & {\footnotesize{}1.08} & {\footnotesize{}1.06} &  & {\footnotesize{}482.16} & {\footnotesize{}1026.68}\tabularnewline
{\footnotesize{}CaTiS$_{3}$} & {\footnotesize{}4.63} & {\footnotesize{}222} & {\footnotesize{}0.78} & {\footnotesize{}0.72} &  & {\footnotesize{}15.10} & {\footnotesize{}13.93} &  & {\footnotesize{}1.15} & {\footnotesize{}1.13} &  & {\footnotesize{}172.02} & {\footnotesize{}221.38}\tabularnewline
{\footnotesize{}CaTiSe$_{3}$} & {\footnotesize{}2.90} & {\footnotesize{}139} & {\footnotesize{}0.43} & {\footnotesize{}0.34} &  & {\footnotesize{}5.19} & {\footnotesize{}4.12} &  & {\footnotesize{}1.08} & {\footnotesize{}1.06} &  & {\footnotesize{}668.29} & {\footnotesize{}1271.42}\tabularnewline
\hline 
\end{tabular}{\footnotesize\par}
\end{table}

Feynman developed an innovative approach to solve the Fr\"ohlich Hamiltonian,
describing the interaction between an electron and a collection of
independent phonon excitations, which behave harmonically and are
incorporated into the quantum field theory \citep{chapter2-23}. As
the electron moves through the lattice, it interacts with the perturbation
it creates, which diminishes exponentially over time. Using Feynman's
method, the effective mass of the polaron ($m_{p}$) for our systems
is also calculated following Eq. \ref{eq:9} and the values are tabulated
in Table \ref{tab:5}. These results reveal that electron-phonon coupling
leads to an increase in the polaron effective mass by 6$\lyxmathsym{\textendash}$63\%,
confirming weak to intermediate carrier-lattice interactions.

Furthermore, the polaron mobility is estimated using the Hellwarth
polaron model through Eq. \ref{eq:10} to confirm the impact of the
increasing polaron effective mass. Polaron mobility quantifies how
easily a polaron moves through a lattice, influenced by its effective
mass and the strength of carrier-lattice interactions. As the polaron's
effective mass increases due to stronger electron-phonon coupling,
the mobility generally decreases, causing the polaron to move more
slowly within the material. This trend is also evident in our systems
of interest (see Table \ref{tab:5}). However, interestingly, mobility
is found to increase with the increase in Ti concentration, and it
boosts significantly in the presence of Se. Notably, these materials
display ambipolar characteristics in polaron mobility, with values
ranging from 17.70 to 668.29 cm$^{2}$V$^{-1}$s$^{-1}$ for electrons
and 34.59 to 1271.42 cm$^{2}$V$^{-1}$s$^{-1}$ for holes. The mobility
values of these alloyed Ti-rich Se-based perovskites are substantially
higher than those found in conventional lead-based HPs (57$-$290
cm$^{2}$V$^{-1}$s$^{-1}$ for electrons and 97$-$230 cm$^{2}$V$^{-1}$s$^{-1}$
for holes, respectively) \citep{chapter2-20,chapter5-15}. Overall,
the polaronic properties suggest that doping at both cationic and
anionic sites enhances photovoltaic performance in these systems.

\subsection{Spectroscopic Limited Maximum Efficiency:}

As discussed, these alloyed perovskites exhibit a high absorption
coefficient and a direct bandgap within the optimal range, making
them promising candidates for efficient solar absorbers. To further
evaluate the photovoltaic performance of these materials, the spectroscopic
limited maximum efficiency (SLME), as introduced by Yu and Zunger
\citep{chapter3-30}, is also calculated (for details, see the Supplemental
Material). SLME is an improved performance metric over the Shockley-Queisser
(SQ) efficiency limit \citep{chapter3-31}, specifically designed
to assess the theoretical maximum efficiency of a thin-film absorber
material. The latter is less realistic as it overlooks losses from
radiative recombination caused by the non-conservation of absorbed
photon momentum. The SLME model considers various factors, including
the magnitude and nature (direct or indirect) of the bandgap, the
shape of the absorption spectrum, the absorber layer thickness, the
material-specific non-radiative recombination losses, and the temperature.
Therefore, the theoretical SLME of our investigated perovskites is
evaluated at 293.15 K using the standard solar spectrum (AM1.5G),
along with the absorption coefficient, material thickness, and the
electronic G$_{0}$W$_{0}$@PBE bandgap as input parameters.

\begin{figure}[H]
\begin{centering}
\includegraphics[width=1\textwidth,height=1\textheight,keepaspectratio]{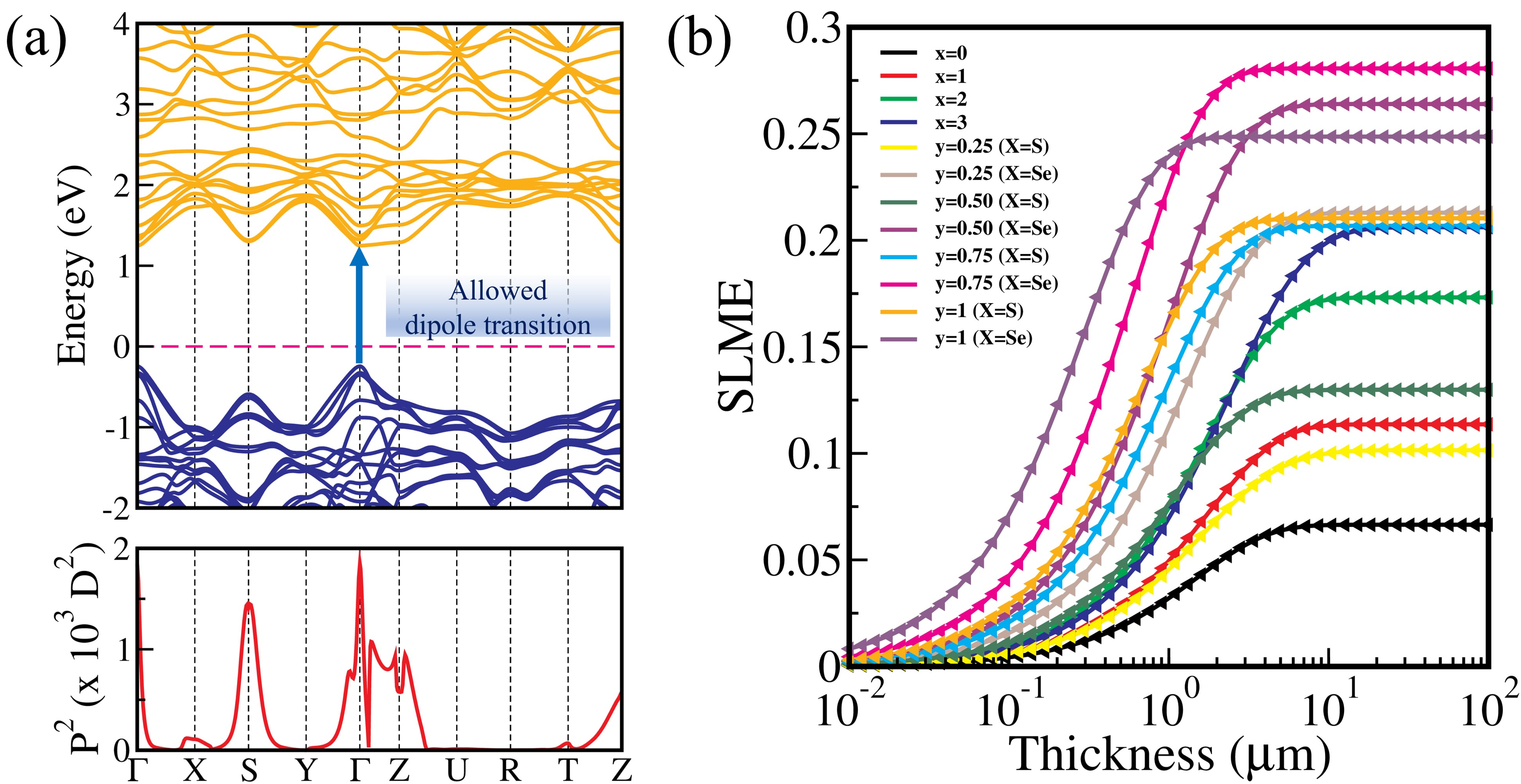}
\par\end{centering}
\caption{\label{fig6}(a) Electronic band structure and transition probability
(square of the transition dipole moment matrix elements) of CaHf$_{0.25}$Ti$_{0.75}$Se$_{3}$
perovskite calculated using G$_{0}$W$_{0}$@PBE method, and (b) spectroscopic
limited maximum efficiency of CaHfS$_{3-x}$Se$_{x}$ ($x$ = 0, 1,
2, 3) and CaHf$_{1-y}$Ti$_{y}$X$_{3}$ ($y$ = 0.25, 0.5, 0.75,
1; X = S, Se) CPs calculated using BSE@G$_{0}$W$_{0}$@PBE method.}

\end{figure}

To perform the SLME calculation, we first assess the optical transition
possibility from VBM to CBM in these compounds. It has been observed
that, despite having a direct electronic bandgap, the optically allowed
dipole transition from VBM to CBM can still be forbidden in some perovskites
\citep{chapter3-19,chapter5-16}. This is due to the inversion symmetry
in these structures, which results in the VBM and CBM having the same
parity. To confirm the possibility of optical transitions from the
VBM to the CBM in our investigated systems, the transition dipole
moment matrix element (P) is computed, since its square (P$^{2}$)
determines the transition probability between the initial (VBM) and
final (CBM) states. The P$^{2}$ values for each system are plotted
and shown in Figure \ref{fig6}(a) and Figure S1, below their respective
G$_{0}$W$_{0}$@PBE band structure plots. These results indicate
that all the compounds exhibit allowed dipole transitions at the $\Gamma$
point. Note that CaTiSe$_{3}$ has an indirect electronic bandgap
but exhibits an optically allowed dipole transition at its lowest
direct band edge. For this indirect bandgap material, non-radiative
recombination significantly influences the SLME calculation. Meanwhile,
radiative recombination is modulated by a factor $f_{r}=e^{(E_{g}-E_{g}^{da})/k_{B}T}$,
where $E_{g}$ is the fundamental bandgap, $E_{g}^{da}$ is the direct
allowed bandgap, $k_{B}$ is the Boltzmann constant, and $T$ is the
temperature \citep{chapter3-30} (for details, see the Supplemental
Material).

Next, the thickness dependence SLME calculations are conducted for
all the investigated compounds using BSE@G$_{0}$W$_{0}$@PBE method,
as shown in Figure \ref{fig6}(b). It is found that the SLME of these
materials increases with thickness up to a certain point, after which
it levels off. The maximum SLME values of these CPs at a 10 $\mu$m
absorber layer thickness are found in the range of 6.65\%$-$28.06\%.
Among these compounds, CaHf$_{0.25}$Ti$_{0.75}$Se$_{3}$ exhibits
the highest SLME of 28.06\% at a thickness of 3 $\mu$m, which is
very close to the maximum SLME of CH$_{3}$NH$_{3}$PbI$_{3}$ (28.97\%
at 2 $\mu$m thickness) \citep{chapter5-13}. Our SLME results demonstrate
that doping at both the cationic and anionic sites substantially improves
the photovoltaic potential of the studied systems. Overall, these
doped CP alloys emerge as more promising candidates for photovoltaic
applications compared to pristine CsPbI$_{3}$, ABX$_{3}$ (A = Ca,
Sr, Ba; B = Zr, Hf, and X= S, Se), and other halide perovskites \citep{chapter3-19,chapter3-33,chapter3-34,chapter3-32,chapter1-63,chapter5-16,chapter5-17}.

\section{Conclusion:}

In this work, we conducted a deep dive into the electronic, optical,
excitonic, and polaronic properties of alloyed chalcogenide perovskites\textemdash CaHfS$_{3-x}$Se$_{x}$
($x$ = 0, 1, 2, 3) and CaHf$_{1-y}$Ti$_{y}$X$_{3}$ ($y$ = 0,
0.25, 0.50, 0.75, 1; X = S, Se)\textemdash using cutting-edge first-principles
based simulations. Phonon dispersion curves and elastic properties
confirm these materials are rock-solid. Our electronic structure analysis
reveals direct G$_{0}$W$_{0}$@PBE bandgaps between 1.29\textendash 2.67
eV, coupled with low carrier effective masses, signaling exceptional
ambipolar mobility. Optical properties, calculated via the Bethe-Salpeter
equation, show an intense absorption onset stretching from the near-infrared
to the visible spectrum, making them perfect for solar harvesting.
Exciton binding energy drops (0.008\textendash 0.204 eV), while exciton
lifetimes extend with Ti and Se alloying\textemdash unlocking more
excellent photophysical performance. Polaron analysis adds another
layer of promise: weak-to-intermediate carrier-phonon coupling, with
charge-carrier mobilities reaching staggering levels\textemdash up
to 668.29 cm$^{2}$V$^{-1}$s$^{-1}$ for electrons and 1271.42 cm$^{2}$V$^{-1}$s$^{-1}$
for holes, for the Ti-rich Se-based perovskites\textemdash far outshining
conventional lead-based and pristine chalcogenide perovskites. Our
theoretical efficiency estimates hit a game-changing 28.06\%, positioning
these materials as frontrunners for lead-free, high-efficiency photovoltaic
devices. Alloyed chalcogenide perovskites may rewrite the rules for
next-gen optoelectronics\textemdash brighter, faster, and more sustainable
than ever before.
\begin{acknowledgments}
The authors would like to acknowledge the Council of Scientific and
Industrial Research (CSIR), Government of India {[}Grant No. 3WS(007)/2023-24/EMR-II/ASPIRE{]}
for financial support. S.A. would like to acknowledge the Council
of Scientific and Industrial Research (CSIR), Government of India
{[}Grant No. 09/1128(11453)/2021-EMR-I{]} for Senior Research Fellowship.
The authors acknowledge the High Performance Computing Cluster (HPCC)
\textquoteleft Magus\textquoteright{} at Shiv Nadar Institution of
Eminence for providing computational resources that have contributed
to the research results reported within this paper.
\end{acknowledgments}

\bibliographystyle{apsrev4-2}
\bibliography{refs}

\begin{thebibliography}{70}%
\makeatletter
\providecommand \@ifxundefined [1]{%
 \@ifx{#1\undefined}
}%
\providecommand \@ifnum [1]{%
 \ifnum #1\expandafter \@firstoftwo
 \else \expandafter \@secondoftwo
 \fi
}%
\providecommand \@ifx [1]{%
 \ifx #1\expandafter \@firstoftwo
 \else \expandafter \@secondoftwo
 \fi
}%
\providecommand \natexlab [1]{#1}%
\providecommand \enquote  [1]{``#1''}%
\providecommand \bibnamefont  [1]{#1}%
\providecommand \bibfnamefont [1]{#1}%
\providecommand \citenamefont [1]{#1}%
\providecommand \href@noop [0]{\@secondoftwo}%
\providecommand \href [0]{\begingroup \@sanitize@url \@href}%
\providecommand \@href[1]{\@@startlink{#1}\@@href}%
\providecommand \@@href[1]{\endgroup#1\@@endlink}%
\providecommand \@sanitize@url [0]{\catcode `\\12\catcode `\$12\catcode `\&12\catcode `\#12\catcode `\^12\catcode `\_12\catcode `\%12\relax}%
\providecommand \@@startlink[1]{}%
\providecommand \@@endlink[0]{}%
\providecommand \url  [0]{\begingroup\@sanitize@url \@url }%
\providecommand \@url [1]{\endgroup\@href {#1}{\urlprefix }}%
\providecommand \urlprefix  [0]{URL }%
\providecommand \Eprint [0]{\href }%
\providecommand \doibase [0]{https://doi.org/}%
\providecommand \selectlanguage [0]{\@gobble}%
\providecommand \bibinfo  [0]{\@secondoftwo}%
\providecommand \bibfield  [0]{\@secondoftwo}%
\providecommand \translation [1]{[#1]}%
\providecommand \BibitemOpen [0]{}%
\providecommand \bibitemStop [0]{}%
\providecommand \bibitemNoStop [0]{.\EOS\space}%
\providecommand \EOS [0]{\spacefactor3000\relax}%
\providecommand \BibitemShut  [1]{\csname bibitem#1\endcsname}%
\let\auto@bib@innerbib\@empty
\bibitem [{\citenamefont {Kojima}\ \emph {et~al.}(2009)\citenamefont {Kojima}, \citenamefont {Teshima}, \citenamefont {Shirai},\ and\ \citenamefont {Miyasaka}}]{chapter2-12}%
  \BibitemOpen
  \bibfield  {author} {\bibinfo {author} {\bibfnamefont {A.}~\bibnamefont {Kojima}}, \bibinfo {author} {\bibfnamefont {K.}~\bibnamefont {Teshima}}, \bibinfo {author} {\bibfnamefont {Y.}~\bibnamefont {Shirai}},\ and\ \bibinfo {author} {\bibfnamefont {T.}~\bibnamefont {Miyasaka}},\ }\href {https://doi.org/10.1021/ja809598r} {\bibfield  {journal} {\bibinfo  {journal} {J. Am. Chem. Soc.}\ }\textbf {\bibinfo {volume} {131}},\ \bibinfo {pages} {6050} (\bibinfo {year} {2009})},\ \Eprint {https://arxiv.org/abs/https://doi.org/10.1021/ja809598r} {https://doi.org/10.1021/ja809598r} \BibitemShut {NoStop}%
\bibitem [{\citenamefont {Park}(2013)}]{chapter3-8}%
  \BibitemOpen
  \bibfield  {author} {\bibinfo {author} {\bibfnamefont {N.-G.}\ \bibnamefont {Park}},\ }\href {https://doi.org/10.1021/jz400892a} {\bibfield  {journal} {\bibinfo  {journal} {J. Phys. Chem. Lett.}\ }\textbf {\bibinfo {volume} {4}},\ \bibinfo {pages} {2423} (\bibinfo {year} {2013})},\ \Eprint {https://arxiv.org/abs/https://doi.org/10.1021/jz400892a} {https://doi.org/10.1021/jz400892a} \BibitemShut {NoStop}%
\bibitem [{\citenamefont {Berry}\ \emph {et~al.}(2015)\citenamefont {Berry}, \citenamefont {Buonassisi}, \citenamefont {Egger}, \citenamefont {Hodes}, \citenamefont {Kronik}, \citenamefont {Loo}, \citenamefont {Lubomirsky}, \citenamefont {Marder}, \citenamefont {Mastai}, \citenamefont {Miller}, \citenamefont {Mitzi}, \citenamefont {Paz}, \citenamefont {Rappe}, \citenamefont {Riess}, \citenamefont {Rybtchinski}, \citenamefont {Stafsudd}, \citenamefont {Stevanovic}, \citenamefont {Toney}, \citenamefont {Zitoun}, \citenamefont {Kahn}, \citenamefont {Ginley},\ and\ \citenamefont {Cahen}}]{chapter3-9}%
  \BibitemOpen
  \bibfield  {author} {\bibinfo {author} {\bibfnamefont {J.}~\bibnamefont {Berry}}, \bibinfo {author} {\bibfnamefont {T.}~\bibnamefont {Buonassisi}}, \bibinfo {author} {\bibfnamefont {D.~A.}\ \bibnamefont {Egger}}, \bibinfo {author} {\bibfnamefont {G.}~\bibnamefont {Hodes}}, \bibinfo {author} {\bibfnamefont {L.}~\bibnamefont {Kronik}}, \bibinfo {author} {\bibfnamefont {Y.-L.}\ \bibnamefont {Loo}}, \bibinfo {author} {\bibfnamefont {I.}~\bibnamefont {Lubomirsky}}, \bibinfo {author} {\bibfnamefont {S.~R.}\ \bibnamefont {Marder}}, \bibinfo {author} {\bibfnamefont {Y.}~\bibnamefont {Mastai}}, \bibinfo {author} {\bibfnamefont {J.~S.}\ \bibnamefont {Miller}}, \bibinfo {author} {\bibfnamefont {D.~B.}\ \bibnamefont {Mitzi}}, \bibinfo {author} {\bibfnamefont {Y.}~\bibnamefont {Paz}}, \bibinfo {author} {\bibfnamefont {A.~M.}\ \bibnamefont {Rappe}}, \bibinfo {author} {\bibfnamefont {I.}~\bibnamefont {Riess}}, \bibinfo {author} {\bibfnamefont {B.}~\bibnamefont {Rybtchinski}}, \bibinfo {author} {\bibfnamefont
  {O.}~\bibnamefont {Stafsudd}}, \bibinfo {author} {\bibfnamefont {V.}~\bibnamefont {Stevanovic}}, \bibinfo {author} {\bibfnamefont {M.~F.}\ \bibnamefont {Toney}}, \bibinfo {author} {\bibfnamefont {D.}~\bibnamefont {Zitoun}}, \bibinfo {author} {\bibfnamefont {A.}~\bibnamefont {Kahn}}, \bibinfo {author} {\bibfnamefont {D.}~\bibnamefont {Ginley}},\ and\ \bibinfo {author} {\bibfnamefont {D.}~\bibnamefont {Cahen}},\ }\href {https://doi.org/https://doi.org/10.1002/adma.201502294} {\bibfield  {journal} {\bibinfo  {journal} {Adv. Mater.}\ }\textbf {\bibinfo {volume} {27}},\ \bibinfo {pages} {5102} (\bibinfo {year} {2015})},\ \Eprint {https://arxiv.org/abs/https://onlinelibrary.wiley.com/doi/pdf/10.1002/adma.201502294} {https://onlinelibrary.wiley.com/doi/pdf/10.1002/adma.201502294} \BibitemShut {NoStop}%
\bibitem [{\citenamefont {Egger}\ \emph {et~al.}(2016)\citenamefont {Egger}, \citenamefont {Rappe},\ and\ \citenamefont {Kronik}}]{chapter3-10}%
  \BibitemOpen
  \bibfield  {author} {\bibinfo {author} {\bibfnamefont {D.~A.}\ \bibnamefont {Egger}}, \bibinfo {author} {\bibfnamefont {A.~M.}\ \bibnamefont {Rappe}},\ and\ \bibinfo {author} {\bibfnamefont {L.}~\bibnamefont {Kronik}},\ }\href {https://doi.org/10.1021/acs.accounts.5b00540} {\bibfield  {journal} {\bibinfo  {journal} {Acc. Chem. Res.}\ }\textbf {\bibinfo {volume} {49}},\ \bibinfo {pages} {573} (\bibinfo {year} {2016})},\ \Eprint {https://arxiv.org/abs/https://doi.org/10.1021/acs.accounts.5b00540} {https://doi.org/10.1021/acs.accounts.5b00540} \BibitemShut {NoStop}%
\bibitem [{\citenamefont {NREL}(2024)}]{chapter2-53}%
  \BibitemOpen
  \bibfield  {author} {\bibinfo {author} {\bibnamefont {NREL}},\ }\href@noop {} {\bibfield  {journal} {\bibinfo  {journal} {Photovoltaic Research}\ } (\bibinfo {year} {2024})}\BibitemShut {NoStop}%
\bibitem [{\citenamefont {Straus}\ \emph {et~al.}(2020)\citenamefont {Straus}, \citenamefont {Guo}, \citenamefont {Abeykoon},\ and\ \citenamefont {Cava}}]{chapter3-12}%
  \BibitemOpen
  \bibfield  {author} {\bibinfo {author} {\bibfnamefont {D.~B.}\ \bibnamefont {Straus}}, \bibinfo {author} {\bibfnamefont {S.}~\bibnamefont {Guo}}, \bibinfo {author} {\bibfnamefont {A.~M.}\ \bibnamefont {Abeykoon}},\ and\ \bibinfo {author} {\bibfnamefont {R.~J.}\ \bibnamefont {Cava}},\ }\href {https://doi.org/https://doi.org/10.1002/adma.202001069} {\bibfield  {journal} {\bibinfo  {journal} {Adv.Mater.}\ }\textbf {\bibinfo {volume} {32}},\ \bibinfo {pages} {2001069} (\bibinfo {year} {2020})},\ \Eprint {https://arxiv.org/abs/https://onlinelibrary.wiley.com/doi/pdf/10.1002/adma.202001069} {https://onlinelibrary.wiley.com/doi/pdf/10.1002/adma.202001069} \BibitemShut {NoStop}%
\bibitem [{\citenamefont {Babayigit}\ \emph {et~al.}(2016)\citenamefont {Babayigit}, \citenamefont {Ethirajan}, \citenamefont {Muller},\ and\ \citenamefont {Conings}}]{chapter1-16}%
  \BibitemOpen
  \bibfield  {author} {\bibinfo {author} {\bibfnamefont {A.}~\bibnamefont {Babayigit}}, \bibinfo {author} {\bibfnamefont {A.}~\bibnamefont {Ethirajan}}, \bibinfo {author} {\bibfnamefont {M.}~\bibnamefont {Muller}},\ and\ \bibinfo {author} {\bibfnamefont {B.}~\bibnamefont {Conings}},\ }\href {https://doi.org/10.1038/nmat4572} {\bibfield  {journal} {\bibinfo  {journal} {Nat. Mater.}\ }\textbf {\bibinfo {volume} {15}},\ \bibinfo {pages} {247} (\bibinfo {year} {2016})}\BibitemShut {NoStop}%
\bibitem [{\citenamefont {Tiwari}\ \emph {et~al.}(2021)\citenamefont {Tiwari}, \citenamefont {Hutter},\ and\ \citenamefont {Longo}}]{chapter3-11}%
  \BibitemOpen
  \bibfield  {author} {\bibinfo {author} {\bibfnamefont {D.}~\bibnamefont {Tiwari}}, \bibinfo {author} {\bibfnamefont {O.~S.}\ \bibnamefont {Hutter}},\ and\ \bibinfo {author} {\bibfnamefont {G.}~\bibnamefont {Longo}},\ }\href {https://doi.org/10.1088/2515-7655/abf41c} {\bibfield  {journal} {\bibinfo  {journal} {J. Phys. Energy}\ }\textbf {\bibinfo {volume} {3}},\ \bibinfo {pages} {034010} (\bibinfo {year} {2021})}\BibitemShut {NoStop}%
\bibitem [{\citenamefont {Niu}\ \emph {et~al.}(2017)\citenamefont {Niu}, \citenamefont {Huyan}, \citenamefont {Liu}, \citenamefont {Yeung}, \citenamefont {Ye}, \citenamefont {Blankemeier}, \citenamefont {Orvis}, \citenamefont {Sarkar}, \citenamefont {Singh}, \citenamefont {Kapadia},\ and\ \citenamefont {Ravichandran}}]{chapter3-16}%
  \BibitemOpen
  \bibfield  {author} {\bibinfo {author} {\bibfnamefont {S.}~\bibnamefont {Niu}}, \bibinfo {author} {\bibfnamefont {H.}~\bibnamefont {Huyan}}, \bibinfo {author} {\bibfnamefont {Y.}~\bibnamefont {Liu}}, \bibinfo {author} {\bibfnamefont {M.}~\bibnamefont {Yeung}}, \bibinfo {author} {\bibfnamefont {K.}~\bibnamefont {Ye}}, \bibinfo {author} {\bibfnamefont {L.}~\bibnamefont {Blankemeier}}, \bibinfo {author} {\bibfnamefont {T.}~\bibnamefont {Orvis}}, \bibinfo {author} {\bibfnamefont {D.}~\bibnamefont {Sarkar}}, \bibinfo {author} {\bibfnamefont {D.~J.}\ \bibnamefont {Singh}}, \bibinfo {author} {\bibfnamefont {R.}~\bibnamefont {Kapadia}},\ and\ \bibinfo {author} {\bibfnamefont {J.}~\bibnamefont {Ravichandran}},\ }\href {https://doi.org/https://doi.org/10.1002/adma.201604733} {\bibfield  {journal} {\bibinfo  {journal} {Adv. Mater.}\ }\textbf {\bibinfo {volume} {29}},\ \bibinfo {pages} {1604733} (\bibinfo {year} {2017})},\ \Eprint {https://arxiv.org/abs/https://onlinelibrary.wiley.com/doi/pdf/10.1002/adma.201604733}
  {https://onlinelibrary.wiley.com/doi/pdf/10.1002/adma.201604733} \BibitemShut {NoStop}%
\bibitem [{\citenamefont {Wu}\ \emph {et~al.}(2021)\citenamefont {Wu}, \citenamefont {Gao}, \citenamefont {Chai}, \citenamefont {Ming}, \citenamefont {Chen}, \citenamefont {Zeng}, \citenamefont {Zhang}, \citenamefont {Zhang},\ and\ \citenamefont {Sun}}]{chapter3-17}%
  \BibitemOpen
  \bibfield  {author} {\bibinfo {author} {\bibfnamefont {X.}~\bibnamefont {Wu}}, \bibinfo {author} {\bibfnamefont {W.}~\bibnamefont {Gao}}, \bibinfo {author} {\bibfnamefont {J.}~\bibnamefont {Chai}}, \bibinfo {author} {\bibfnamefont {C.}~\bibnamefont {Ming}}, \bibinfo {author} {\bibfnamefont {M.}~\bibnamefont {Chen}}, \bibinfo {author} {\bibfnamefont {H.}~\bibnamefont {Zeng}}, \bibinfo {author} {\bibfnamefont {P.}~\bibnamefont {Zhang}}, \bibinfo {author} {\bibfnamefont {S.}~\bibnamefont {Zhang}},\ and\ \bibinfo {author} {\bibfnamefont {Y.-Y.}\ \bibnamefont {Sun}},\ }\href {https://doi.org/10.1007/s40843-021-1683-0} {\bibfield  {journal} {\bibinfo  {journal} {Sci. China Mater.}\ }\textbf {\bibinfo {volume} {64}},\ \bibinfo {pages} {2976} (\bibinfo {year} {2021})}\BibitemShut {NoStop}%
\bibitem [{\citenamefont {Sun}\ \emph {et~al.}(2015)\citenamefont {Sun}, \citenamefont {Agiorgousis}, \citenamefont {Zhang},\ and\ \citenamefont {Zhang}}]{chapter3-18}%
  \BibitemOpen
  \bibfield  {author} {\bibinfo {author} {\bibfnamefont {Y.-Y.}\ \bibnamefont {Sun}}, \bibinfo {author} {\bibfnamefont {M.~L.}\ \bibnamefont {Agiorgousis}}, \bibinfo {author} {\bibfnamefont {P.}~\bibnamefont {Zhang}},\ and\ \bibinfo {author} {\bibfnamefont {S.}~\bibnamefont {Zhang}},\ }\href {https://doi.org/10.1021/nl504046x} {\bibfield  {journal} {\bibinfo  {journal} {Nano Lett.}\ }\textbf {\bibinfo {volume} {15}},\ \bibinfo {pages} {581} (\bibinfo {year} {2015})},\ \Eprint {https://arxiv.org/abs/https://doi.org/10.1021/nl504046x} {https://doi.org/10.1021/nl504046x} \BibitemShut {NoStop}%
\bibitem [{\citenamefont {Liu}\ \emph {et~al.}(2023)\citenamefont {Liu}, \citenamefont {Zeng}, \citenamefont {Peng},\ and\ \citenamefont {Sa}}]{chapter3-3}%
  \BibitemOpen
  \bibfield  {author} {\bibinfo {author} {\bibfnamefont {D.}~\bibnamefont {Liu}}, \bibinfo {author} {\bibfnamefont {H.}~\bibnamefont {Zeng}}, \bibinfo {author} {\bibfnamefont {H.}~\bibnamefont {Peng}},\ and\ \bibinfo {author} {\bibfnamefont {R.}~\bibnamefont {Sa}},\ }\href {https://doi.org/10.1039/D3CP01522J} {\bibfield  {journal} {\bibinfo  {journal} {Phys. Chem. Chem. Phys.}\ }\textbf {\bibinfo {volume} {25}},\ \bibinfo {pages} {13755} (\bibinfo {year} {2023})}\BibitemShut {NoStop}%
\bibitem [{\citenamefont {Swarnkar}\ \emph {et~al.}(2019)\citenamefont {Swarnkar}, \citenamefont {Mir}, \citenamefont {Chakraborty}, \citenamefont {Jagadeeswararao}, \citenamefont {Sheikh},\ and\ \citenamefont {Nag}}]{chapter3-13}%
  \BibitemOpen
  \bibfield  {author} {\bibinfo {author} {\bibfnamefont {A.}~\bibnamefont {Swarnkar}}, \bibinfo {author} {\bibfnamefont {W.~J.}\ \bibnamefont {Mir}}, \bibinfo {author} {\bibfnamefont {R.}~\bibnamefont {Chakraborty}}, \bibinfo {author} {\bibfnamefont {M.}~\bibnamefont {Jagadeeswararao}}, \bibinfo {author} {\bibfnamefont {T.}~\bibnamefont {Sheikh}},\ and\ \bibinfo {author} {\bibfnamefont {A.}~\bibnamefont {Nag}},\ }\href {https://doi.org/10.1021/acs.chemmater.8b04178} {\bibfield  {journal} {\bibinfo  {journal} {Chem. Mater.}\ }\textbf {\bibinfo {volume} {31}},\ \bibinfo {pages} {565} (\bibinfo {year} {2019})},\ \Eprint {https://arxiv.org/abs/https://doi.org/10.1021/acs.chemmater.8b04178} {https://doi.org/10.1021/acs.chemmater.8b04178} \BibitemShut {NoStop}%
\bibitem [{\citenamefont {Basera}\ and\ \citenamefont {Bhattacharya}(2022)}]{chapter3-19}%
  \BibitemOpen
  \bibfield  {author} {\bibinfo {author} {\bibfnamefont {P.}~\bibnamefont {Basera}}\ and\ \bibinfo {author} {\bibfnamefont {S.}~\bibnamefont {Bhattacharya}},\ }\href {https://doi.org/10.1021/acs.jpclett.2c01337} {\bibfield  {journal} {\bibinfo  {journal} {J. Phys. Chem. Lett.}\ }\textbf {\bibinfo {volume} {13}},\ \bibinfo {pages} {6439} (\bibinfo {year} {2022})},\ \Eprint {https://arxiv.org/abs/https://doi.org/10.1021/acs.jpclett.2c01337} {https://doi.org/10.1021/acs.jpclett.2c01337} \BibitemShut {NoStop}%
\bibitem [{\citenamefont {Lelieveld}\ and\ \citenamefont {IJdo}(1980)}]{chapter3-37}%
  \BibitemOpen
  \bibfield  {author} {\bibinfo {author} {\bibfnamefont {R.}~\bibnamefont {Lelieveld}}\ and\ \bibinfo {author} {\bibfnamefont {D.~J.~W.}\ \bibnamefont {IJdo}},\ }\href {https://doi.org/10.1107/S056774088000845X} {\bibfield  {journal} {\bibinfo  {journal} {Acta Cryst. B}\ }\textbf {\bibinfo {volume} {36}},\ \bibinfo {pages} {2223} (\bibinfo {year} {1980})}\BibitemShut {NoStop}%
\bibitem [{\citenamefont {Fix}\ \emph {et~al.}(2023)\citenamefont {Fix}, \citenamefont {Raissi}, \citenamefont {Muller}, \citenamefont {Bouillet}, \citenamefont {Preziosi},\ and\ \citenamefont {Slaoui}}]{chapter6-1}%
  \BibitemOpen
  \bibfield  {author} {\bibinfo {author} {\bibfnamefont {T.}~\bibnamefont {Fix}}, \bibinfo {author} {\bibfnamefont {S.}~\bibnamefont {Raissi}}, \bibinfo {author} {\bibfnamefont {D.}~\bibnamefont {Muller}}, \bibinfo {author} {\bibfnamefont {C.}~\bibnamefont {Bouillet}}, \bibinfo {author} {\bibfnamefont {D.}~\bibnamefont {Preziosi}},\ and\ \bibinfo {author} {\bibfnamefont {A.}~\bibnamefont {Slaoui}},\ }\href {https://doi.org/https://doi.org/10.1016/j.jallcom.2023.171272} {\bibfield  {journal} {\bibinfo  {journal} {J. Alloys Compd.}\ }\textbf {\bibinfo {volume} {964}},\ \bibinfo {pages} {171272} (\bibinfo {year} {2023})}\BibitemShut {NoStop}%
\bibitem [{\citenamefont {Kumar}\ \emph {et~al.}(2021)\citenamefont {Kumar}, \citenamefont {Singh}, \citenamefont {Gill},\ and\ \citenamefont {Bhattacharya}}]{chapter1-63}%
  \BibitemOpen
  \bibfield  {author} {\bibinfo {author} {\bibfnamefont {M.}~\bibnamefont {Kumar}}, \bibinfo {author} {\bibfnamefont {A.}~\bibnamefont {Singh}}, \bibinfo {author} {\bibfnamefont {D.}~\bibnamefont {Gill}},\ and\ \bibinfo {author} {\bibfnamefont {S.}~\bibnamefont {Bhattacharya}},\ }\href {https://doi.org/10.1021/acs.jpclett.1c01034} {\bibfield  {journal} {\bibinfo  {journal} {J. Phys. Chem. Lett.}\ }\textbf {\bibinfo {volume} {12}},\ \bibinfo {pages} {5301} (\bibinfo {year} {2021})}\BibitemShut {NoStop}%
\bibitem [{\citenamefont {Adhikari}\ and\ \citenamefont {Johari}(2024{\natexlab{a}})}]{chapter5-16}%
  \BibitemOpen
  \bibfield  {author} {\bibinfo {author} {\bibfnamefont {S.}~\bibnamefont {Adhikari}}\ and\ \bibinfo {author} {\bibfnamefont {P.}~\bibnamefont {Johari}},\ }\href {https://doi.org/10.1103/PhysRevB.109.174114} {\bibfield  {journal} {\bibinfo  {journal} {Phys. Rev. B}\ }\textbf {\bibinfo {volume} {109}},\ \bibinfo {pages} {174114} (\bibinfo {year} {2024}{\natexlab{a}})}\BibitemShut {NoStop}%
\bibitem [{\citenamefont {Ghaithan}\ \emph {et~al.}(2021)\citenamefont {Ghaithan}, \citenamefont {Qaid}, \citenamefont {Alahmed}, \citenamefont {Hezam}, \citenamefont {Lyras}, \citenamefont {Amer},\ and\ \citenamefont {Aldwayyan}}]{chapter5-11}%
  \BibitemOpen
  \bibfield  {author} {\bibinfo {author} {\bibfnamefont {H.~M.}\ \bibnamefont {Ghaithan}}, \bibinfo {author} {\bibfnamefont {S.~M.~H.}\ \bibnamefont {Qaid}}, \bibinfo {author} {\bibfnamefont {Z.~A.}\ \bibnamefont {Alahmed}}, \bibinfo {author} {\bibfnamefont {M.}~\bibnamefont {Hezam}}, \bibinfo {author} {\bibfnamefont {A.}~\bibnamefont {Lyras}}, \bibinfo {author} {\bibfnamefont {M.}~\bibnamefont {Amer}},\ and\ \bibinfo {author} {\bibfnamefont {A.~S.}\ \bibnamefont {Aldwayyan}},\ }\href {https://doi.org/10.1021/acs.jpcc.0c07983} {\bibfield  {journal} {\bibinfo  {journal} {J. Phys. Chem. C}\ }\textbf {\bibinfo {volume} {125}},\ \bibinfo {pages} {886} (\bibinfo {year} {2021})},\ \Eprint {https://arxiv.org/abs/https://doi.org/10.1021/acs.jpcc.0c07983} {https://doi.org/10.1021/acs.jpcc.0c07983} \BibitemShut {NoStop}%
\bibitem [{\citenamefont {Comin}\ \emph {et~al.}(2015)\citenamefont {Comin}, \citenamefont {Walters}, \citenamefont {Thibau}, \citenamefont {Voznyy}, \citenamefont {Lu},\ and\ \citenamefont {Sargent}}]{chapter5-12}%
  \BibitemOpen
  \bibfield  {author} {\bibinfo {author} {\bibfnamefont {R.}~\bibnamefont {Comin}}, \bibinfo {author} {\bibfnamefont {G.}~\bibnamefont {Walters}}, \bibinfo {author} {\bibfnamefont {E.~S.}\ \bibnamefont {Thibau}}, \bibinfo {author} {\bibfnamefont {O.}~\bibnamefont {Voznyy}}, \bibinfo {author} {\bibfnamefont {Z.-H.}\ \bibnamefont {Lu}},\ and\ \bibinfo {author} {\bibfnamefont {E.~H.}\ \bibnamefont {Sargent}},\ }\href {https://doi.org/10.1039/C5TC01718A} {\bibfield  {journal} {\bibinfo  {journal} {J. Mater. Chem. C}\ }\textbf {\bibinfo {volume} {3}},\ \bibinfo {pages} {8839} (\bibinfo {year} {2015})}\BibitemShut {NoStop}%
\bibitem [{\citenamefont {Chen}\ \emph {et~al.}(2018)\citenamefont {Chen}, \citenamefont {Lu}, \citenamefont {Yang},\ and\ \citenamefont {Beard}}]{chapter5-14}%
  \BibitemOpen
  \bibfield  {author} {\bibinfo {author} {\bibfnamefont {X.}~\bibnamefont {Chen}}, \bibinfo {author} {\bibfnamefont {H.}~\bibnamefont {Lu}}, \bibinfo {author} {\bibfnamefont {Y.}~\bibnamefont {Yang}},\ and\ \bibinfo {author} {\bibfnamefont {M.~C.}\ \bibnamefont {Beard}},\ }\href {https://doi.org/10.1021/acs.jpclett.8b00526} {\bibfield  {journal} {\bibinfo  {journal} {J. Phys. Chem. Lett.}\ }\textbf {\bibinfo {volume} {9}},\ \bibinfo {pages} {2595} (\bibinfo {year} {2018})},\ \bibinfo {note} {pMID: 29714488},\ \Eprint {https://arxiv.org/abs/https://doi.org/10.1021/acs.jpclett.8b00526} {https://doi.org/10.1021/acs.jpclett.8b00526} \BibitemShut {NoStop}%
\bibitem [{\citenamefont {Sun}\ \emph {et~al.}(2019)\citenamefont {Sun}, \citenamefont {Chen},\ and\ \citenamefont {Yin}}]{chapter5-13}%
  \BibitemOpen
  \bibfield  {author} {\bibinfo {author} {\bibfnamefont {Q.}~\bibnamefont {Sun}}, \bibinfo {author} {\bibfnamefont {H.}~\bibnamefont {Chen}},\ and\ \bibinfo {author} {\bibfnamefont {W.-J.}\ \bibnamefont {Yin}},\ }\href {https://doi.org/10.1021/acs.chemmater.8b04320} {\bibfield  {journal} {\bibinfo  {journal} {Chem. Mater.}\ }\textbf {\bibinfo {volume} {31}},\ \bibinfo {pages} {244} (\bibinfo {year} {2019})},\ \Eprint {https://arxiv.org/abs/https://doi.org/10.1021/acs.chemmater.8b04320} {https://doi.org/10.1021/acs.chemmater.8b04320} \BibitemShut {NoStop}%
\bibitem [{\citenamefont {Kang}\ and\ \citenamefont {Han}(2018)}]{chapter5-15}%
  \BibitemOpen
  \bibfield  {author} {\bibinfo {author} {\bibfnamefont {Y.}~\bibnamefont {Kang}}\ and\ \bibinfo {author} {\bibfnamefont {S.}~\bibnamefont {Han}},\ }\href {https://doi.org/10.1103/PhysRevApplied.10.044013} {\bibfield  {journal} {\bibinfo  {journal} {Phys. Rev. Appl.}\ }\textbf {\bibinfo {volume} {10}},\ \bibinfo {pages} {044013} (\bibinfo {year} {2018})}\BibitemShut {NoStop}%
\bibitem [{\citenamefont {Zou}\ \emph {et~al.}(2023)\citenamefont {Zou}, \citenamefont {Duan}, \citenamefont {Yang}, \citenamefont {Xu}, \citenamefont {Yang}, \citenamefont {Cui}, \citenamefont {Zhou}, \citenamefont {Wu}, \citenamefont {Wang}, \citenamefont {Lei}, \citenamefont {Zhang},\ and\ \citenamefont {Liu}}]{chapter5-17}%
  \BibitemOpen
  \bibfield  {author} {\bibinfo {author} {\bibfnamefont {H.}~\bibnamefont {Zou}}, \bibinfo {author} {\bibfnamefont {Y.}~\bibnamefont {Duan}}, \bibinfo {author} {\bibfnamefont {S.}~\bibnamefont {Yang}}, \bibinfo {author} {\bibfnamefont {D.}~\bibnamefont {Xu}}, \bibinfo {author} {\bibfnamefont {L.}~\bibnamefont {Yang}}, \bibinfo {author} {\bibfnamefont {J.}~\bibnamefont {Cui}}, \bibinfo {author} {\bibfnamefont {H.}~\bibnamefont {Zhou}}, \bibinfo {author} {\bibfnamefont {M.}~\bibnamefont {Wu}}, \bibinfo {author} {\bibfnamefont {J.}~\bibnamefont {Wang}}, \bibinfo {author} {\bibfnamefont {X.}~\bibnamefont {Lei}}, \bibinfo {author} {\bibfnamefont {N.}~\bibnamefont {Zhang}},\ and\ \bibinfo {author} {\bibfnamefont {Z.}~\bibnamefont {Liu}},\ }\href {https://doi.org/https://doi.org/10.1002/smll.202206205} {\bibfield  {journal} {\bibinfo  {journal} {Small}\ }\textbf {\bibinfo {volume} {19}},\ \bibinfo {pages} {2206205} (\bibinfo {year} {2023})},\ \Eprint
  {https://arxiv.org/abs/https://onlinelibrary.wiley.com/doi/pdf/10.1002/smll.202206205} {https://onlinelibrary.wiley.com/doi/pdf/10.1002/smll.202206205} \BibitemShut {NoStop}%
\bibitem [{\citenamefont {Frost}(2017)}]{chapter2-20}%
  \BibitemOpen
  \bibfield  {author} {\bibinfo {author} {\bibfnamefont {J.~M.}\ \bibnamefont {Frost}},\ }\href {https://doi.org/10.1103/PhysRevB.96.195202} {\bibfield  {journal} {\bibinfo  {journal} {Phys. Rev. B}\ }\textbf {\bibinfo {volume} {96}},\ \bibinfo {pages} {195202} (\bibinfo {year} {2017})}\BibitemShut {NoStop}%
\bibitem [{\citenamefont {Meng}\ \emph {et~al.}(2016)\citenamefont {Meng}, \citenamefont {Saparov}, \citenamefont {Hong}, \citenamefont {Wang}, \citenamefont {Mitzi},\ and\ \citenamefont {Yan}}]{chapter6-4}%
  \BibitemOpen
  \bibfield  {author} {\bibinfo {author} {\bibfnamefont {W.}~\bibnamefont {Meng}}, \bibinfo {author} {\bibfnamefont {B.}~\bibnamefont {Saparov}}, \bibinfo {author} {\bibfnamefont {F.}~\bibnamefont {Hong}}, \bibinfo {author} {\bibfnamefont {J.}~\bibnamefont {Wang}}, \bibinfo {author} {\bibfnamefont {D.~B.}\ \bibnamefont {Mitzi}},\ and\ \bibinfo {author} {\bibfnamefont {Y.}~\bibnamefont {Yan}},\ }\href {https://doi.org/10.1021/acs.chemmater.5b04213} {\bibfield  {journal} {\bibinfo  {journal} {Chem. Mater.}\ }\textbf {\bibinfo {volume} {28}},\ \bibinfo {pages} {821} (\bibinfo {year} {2016})},\ \Eprint {https://arxiv.org/abs/https://doi.org/10.1021/acs.chemmater.5b04213} {https://doi.org/10.1021/acs.chemmater.5b04213} \BibitemShut {NoStop}%
\bibitem [{\citenamefont {Kong}\ \emph {et~al.}(2024)\citenamefont {Kong}, \citenamefont {Dong}, \citenamefont {Yu}, \citenamefont {Guo}, \citenamefont {Cao}, \citenamefont {Chen}, \citenamefont {Ke}, \citenamefont {Zhou}, \citenamefont {Deng}, \citenamefont {Yang},\ and\ \citenamefont {Zhang}}]{chapter6-3}%
  \BibitemOpen
  \bibfield  {author} {\bibinfo {author} {\bibfnamefont {S.}~\bibnamefont {Kong}}, \bibinfo {author} {\bibfnamefont {H.}~\bibnamefont {Dong}}, \bibinfo {author} {\bibfnamefont {Z.}~\bibnamefont {Yu}}, \bibinfo {author} {\bibfnamefont {J.}~\bibnamefont {Guo}}, \bibinfo {author} {\bibfnamefont {K.}~\bibnamefont {Cao}}, \bibinfo {author} {\bibfnamefont {K.}~\bibnamefont {Chen}}, \bibinfo {author} {\bibfnamefont {X.}~\bibnamefont {Ke}}, \bibinfo {author} {\bibfnamefont {C.}~\bibnamefont {Zhou}}, \bibinfo {author} {\bibfnamefont {J.}~\bibnamefont {Deng}}, \bibinfo {author} {\bibfnamefont {S.}~\bibnamefont {Yang}},\ and\ \bibinfo {author} {\bibfnamefont {Y.}~\bibnamefont {Zhang}},\ }\href {https://doi.org/https://doi.org/10.1016/j.ceramint.2023.12.405} {\bibfield  {journal} {\bibinfo  {journal} {Ceram. Int.}\ }\textbf {\bibinfo {volume} {50}},\ \bibinfo {pages} {10889} (\bibinfo {year} {2024})}\BibitemShut {NoStop}%
\bibitem [{\citenamefont {Chami}\ \emph {et~al.}(2021)\citenamefont {Chami}, \citenamefont {Lekdadri}, \citenamefont {Omari}, \citenamefont {Hlil},\ and\ \citenamefont {Chafi}}]{chapter6-6}%
  \BibitemOpen
  \bibfield  {author} {\bibinfo {author} {\bibfnamefont {R.}~\bibnamefont {Chami}}, \bibinfo {author} {\bibfnamefont {A.}~\bibnamefont {Lekdadri}}, \bibinfo {author} {\bibfnamefont {L.}~\bibnamefont {Omari}}, \bibinfo {author} {\bibfnamefont {E.}~\bibnamefont {Hlil}},\ and\ \bibinfo {author} {\bibfnamefont {M.}~\bibnamefont {Chafi}},\ }\href {https://doi.org/https://doi.org/10.1016/j.mtener.2021.100689} {\bibfield  {journal} {\bibinfo  {journal} {Mater. Today Energy}\ }\textbf {\bibinfo {volume} {20}},\ \bibinfo {pages} {100689} (\bibinfo {year} {2021})}\BibitemShut {NoStop}%
\bibitem [{\citenamefont {Chami}\ \emph {et~al.}(2023)\citenamefont {Chami}, \citenamefont {Lekdadri}, \citenamefont {Chafi}, \citenamefont {Omari},\ and\ \citenamefont {Hlil}}]{chapter6-5}%
  \BibitemOpen
  \bibfield  {author} {\bibinfo {author} {\bibfnamefont {R.}~\bibnamefont {Chami}}, \bibinfo {author} {\bibfnamefont {A.}~\bibnamefont {Lekdadri}}, \bibinfo {author} {\bibfnamefont {M.}~\bibnamefont {Chafi}}, \bibinfo {author} {\bibfnamefont {L.}~\bibnamefont {Omari}},\ and\ \bibinfo {author} {\bibfnamefont {E.}~\bibnamefont {Hlil}},\ }\href {https://doi.org/https://doi.org/10.1016/j.ssc.2023.115212} {\bibfield  {journal} {\bibinfo  {journal} {Solid State Commun.}\ }\textbf {\bibinfo {volume} {369}},\ \bibinfo {pages} {115212} (\bibinfo {year} {2023})}\BibitemShut {NoStop}%
\bibitem [{\citenamefont {Liu}\ \emph {et~al.}(2024)\citenamefont {Liu}, \citenamefont {Peng}, \citenamefont {He},\ and\ \citenamefont {Sa}}]{chapter6-7}%
  \BibitemOpen
  \bibfield  {author} {\bibinfo {author} {\bibfnamefont {D.}~\bibnamefont {Liu}}, \bibinfo {author} {\bibfnamefont {H.}~\bibnamefont {Peng}}, \bibinfo {author} {\bibfnamefont {J.}~\bibnamefont {He}},\ and\ \bibinfo {author} {\bibfnamefont {R.}~\bibnamefont {Sa}},\ }\href {https://doi.org/https://doi.org/10.1016/j.mssp.2023.107919} {\bibfield  {journal} {\bibinfo  {journal} {Mater. Sci. Semicond. Process.}\ }\textbf {\bibinfo {volume} {169}},\ \bibinfo {pages} {107919} (\bibinfo {year} {2024})}\BibitemShut {NoStop}%
\bibitem [{\citenamefont {Hohenberg}\ and\ \citenamefont {Kohn}(1964)}]{chapter2-36}%
  \BibitemOpen
  \bibfield  {author} {\bibinfo {author} {\bibfnamefont {P.}~\bibnamefont {Hohenberg}}\ and\ \bibinfo {author} {\bibfnamefont {W.}~\bibnamefont {Kohn}},\ }\href {https://doi.org/10.1103/PhysRev.136.B864} {\bibfield  {journal} {\bibinfo  {journal} {Phys. Rev.}\ }\textbf {\bibinfo {volume} {136}},\ \bibinfo {pages} {B864} (\bibinfo {year} {1964})}\BibitemShut {NoStop}%
\bibitem [{\citenamefont {Kohn}\ and\ \citenamefont {Sham}(1965)}]{chapter2-37}%
  \BibitemOpen
  \bibfield  {author} {\bibinfo {author} {\bibfnamefont {W.}~\bibnamefont {Kohn}}\ and\ \bibinfo {author} {\bibfnamefont {L.~J.}\ \bibnamefont {Sham}},\ }\href {https://doi.org/10.1103/PhysRev.140.A1133} {\bibfield  {journal} {\bibinfo  {journal} {Phys. Rev.}\ }\textbf {\bibinfo {volume} {140}},\ \bibinfo {pages} {A1133} (\bibinfo {year} {1965})}\BibitemShut {NoStop}%
\bibitem [{\citenamefont {Jiang}\ \emph {et~al.}(2012)\citenamefont {Jiang}, \citenamefont {Rinke},\ and\ \citenamefont {Scheffler}}]{chapter3-1}%
  \BibitemOpen
  \bibfield  {author} {\bibinfo {author} {\bibfnamefont {H.}~\bibnamefont {Jiang}}, \bibinfo {author} {\bibfnamefont {P.}~\bibnamefont {Rinke}},\ and\ \bibinfo {author} {\bibfnamefont {M.}~\bibnamefont {Scheffler}},\ }\href {https://doi.org/10.1103/PhysRevB.86.125115} {\bibfield  {journal} {\bibinfo  {journal} {Phys. Rev. B}\ }\textbf {\bibinfo {volume} {86}},\ \bibinfo {pages} {125115} (\bibinfo {year} {2012})}\BibitemShut {NoStop}%
\bibitem [{\citenamefont {Fuchs}\ \emph {et~al.}(2008)\citenamefont {Fuchs}, \citenamefont {R\"odl}, \citenamefont {Schleife},\ and\ \citenamefont {Bechstedt}}]{chapter3-2}%
  \BibitemOpen
  \bibfield  {author} {\bibinfo {author} {\bibfnamefont {F.}~\bibnamefont {Fuchs}}, \bibinfo {author} {\bibfnamefont {C.}~\bibnamefont {R\"odl}}, \bibinfo {author} {\bibfnamefont {A.}~\bibnamefont {Schleife}},\ and\ \bibinfo {author} {\bibfnamefont {F.}~\bibnamefont {Bechstedt}},\ }\href {https://doi.org/10.1103/PhysRevB.78.085103} {\bibfield  {journal} {\bibinfo  {journal} {Phys. Rev. B}\ }\textbf {\bibinfo {volume} {78}},\ \bibinfo {pages} {085103} (\bibinfo {year} {2008})}\BibitemShut {NoStop}%
\bibitem [{\citenamefont {Hedin}(1965)}]{chapter1-69}%
  \BibitemOpen
  \bibfield  {author} {\bibinfo {author} {\bibfnamefont {L.}~\bibnamefont {Hedin}},\ }\href {https://doi.org/10.1103/PhysRev.139.A796} {\bibfield  {journal} {\bibinfo  {journal} {Phys. Rev.}\ }\textbf {\bibinfo {volume} {139}},\ \bibinfo {pages} {A796} (\bibinfo {year} {1965})}\BibitemShut {NoStop}%
\bibitem [{\citenamefont {Hybertsen}\ and\ \citenamefont {Louie}(1985)}]{chapter1-70}%
  \BibitemOpen
  \bibfield  {author} {\bibinfo {author} {\bibfnamefont {M.~S.}\ \bibnamefont {Hybertsen}}\ and\ \bibinfo {author} {\bibfnamefont {S.~G.}\ \bibnamefont {Louie}},\ }\href {https://doi.org/10.1103/PhysRevLett.55.1418} {\bibfield  {journal} {\bibinfo  {journal} {Phys. Rev. Lett.}\ }\textbf {\bibinfo {volume} {55}},\ \bibinfo {pages} {1418} (\bibinfo {year} {1985})}\BibitemShut {NoStop}%
\bibitem [{\citenamefont {Albrecht}\ \emph {et~al.}(1998)\citenamefont {Albrecht}, \citenamefont {Reining}, \citenamefont {Del~Sole},\ and\ \citenamefont {Onida}}]{chapter1-67}%
  \BibitemOpen
  \bibfield  {author} {\bibinfo {author} {\bibfnamefont {S.}~\bibnamefont {Albrecht}}, \bibinfo {author} {\bibfnamefont {L.}~\bibnamefont {Reining}}, \bibinfo {author} {\bibfnamefont {R.}~\bibnamefont {Del~Sole}},\ and\ \bibinfo {author} {\bibfnamefont {G.}~\bibnamefont {Onida}},\ }\href {https://doi.org/10.1103/PhysRevLett.80.4510} {\bibfield  {journal} {\bibinfo  {journal} {Phys. Rev. Lett.}\ }\textbf {\bibinfo {volume} {80}},\ \bibinfo {pages} {4510} (\bibinfo {year} {1998})}\BibitemShut {NoStop}%
\bibitem [{\citenamefont {Rohlfing}\ and\ \citenamefont {Louie}(1998)}]{chapter1-68}%
  \BibitemOpen
  \bibfield  {author} {\bibinfo {author} {\bibfnamefont {M.}~\bibnamefont {Rohlfing}}\ and\ \bibinfo {author} {\bibfnamefont {S.~G.}\ \bibnamefont {Louie}},\ }\href {https://doi.org/10.1103/PhysRevLett.81.2312} {\bibfield  {journal} {\bibinfo  {journal} {Phys. Rev. Lett.}\ }\textbf {\bibinfo {volume} {81}},\ \bibinfo {pages} {2312} (\bibinfo {year} {1998})}\BibitemShut {NoStop}%
\bibitem [{\citenamefont {Gajdo\v{s}}\ \emph {et~al.}(2006)\citenamefont {Gajdo\v{s}}, \citenamefont {Hummer}, \citenamefont {Kresse}, \citenamefont {Furthm\"uller},\ and\ \citenamefont {Bechstedt}}]{chapter1-60}%
  \BibitemOpen
  \bibfield  {author} {\bibinfo {author} {\bibfnamefont {M.}~\bibnamefont {Gajdo\v{s}}}, \bibinfo {author} {\bibfnamefont {K.}~\bibnamefont {Hummer}}, \bibinfo {author} {\bibfnamefont {G.}~\bibnamefont {Kresse}}, \bibinfo {author} {\bibfnamefont {J.}~\bibnamefont {Furthm\"uller}},\ and\ \bibinfo {author} {\bibfnamefont {F.}~\bibnamefont {Bechstedt}},\ }\href {https://doi.org/10.1103/PhysRevB.73.045112} {\bibfield  {journal} {\bibinfo  {journal} {Phys. Rev. B}\ }\textbf {\bibinfo {volume} {73}},\ \bibinfo {pages} {045112} (\bibinfo {year} {2006})}\BibitemShut {NoStop}%
\bibitem [{\citenamefont {Kresse}\ and\ \citenamefont {Furthm\"uller}(1996)}]{chapter1-31}%
  \BibitemOpen
  \bibfield  {author} {\bibinfo {author} {\bibfnamefont {G.}~\bibnamefont {Kresse}}\ and\ \bibinfo {author} {\bibfnamefont {J.}~\bibnamefont {Furthm\"uller}},\ }\href {https://doi.org/10.1103/PhysRevB.54.11169} {\bibfield  {journal} {\bibinfo  {journal} {Phys. Rev. B}\ }\textbf {\bibinfo {volume} {54}},\ \bibinfo {pages} {11169} (\bibinfo {year} {1996})}\BibitemShut {NoStop}%
\bibitem [{\citenamefont {Kresse}\ and\ \citenamefont {Furthm$\ddot{u}$ller}(1996)}]{chapter1-32}%
  \BibitemOpen
  \bibfield  {author} {\bibinfo {author} {\bibfnamefont {G.}~\bibnamefont {Kresse}}\ and\ \bibinfo {author} {\bibfnamefont {J.}~\bibnamefont {Furthm$\ddot{u}$ller}},\ }\href {https://doi.org/https://doi.org/10.1016/0927-0256(96)00008-0} {\bibfield  {journal} {\bibinfo  {journal} {Comput. Mater. Sci.}\ }\textbf {\bibinfo {volume} {6}},\ \bibinfo {pages} {15} (\bibinfo {year} {1996})}\BibitemShut {NoStop}%
\bibitem [{\citenamefont {Bl\"ochl}(1994)}]{chapter1-33}%
  \BibitemOpen
  \bibfield  {author} {\bibinfo {author} {\bibfnamefont {P.~E.}\ \bibnamefont {Bl\"ochl}},\ }\href {https://doi.org/10.1103/PhysRevB.50.17953} {\bibfield  {journal} {\bibinfo  {journal} {Phys. Rev. B}\ }\textbf {\bibinfo {volume} {50}},\ \bibinfo {pages} {17953} (\bibinfo {year} {1994})}\BibitemShut {NoStop}%
\bibitem [{\citenamefont {Perdew}\ \emph {et~al.}(1996)\citenamefont {Perdew}, \citenamefont {Burke},\ and\ \citenamefont {Ernzerhof}}]{chapter1-34}%
  \BibitemOpen
  \bibfield  {author} {\bibinfo {author} {\bibfnamefont {J.~P.}\ \bibnamefont {Perdew}}, \bibinfo {author} {\bibfnamefont {K.}~\bibnamefont {Burke}},\ and\ \bibinfo {author} {\bibfnamefont {M.}~\bibnamefont {Ernzerhof}},\ }\href {https://doi.org/10.1103/PhysRevLett.77.3865} {\bibfield  {journal} {\bibinfo  {journal} {Phys. Rev. Lett.}\ }\textbf {\bibinfo {volume} {77}},\ \bibinfo {pages} {3865} (\bibinfo {year} {1996})}\BibitemShut {NoStop}%
\bibitem [{\citenamefont {Momma}\ and\ \citenamefont {Izumi}(2011)}]{chapter2-3}%
  \BibitemOpen
  \bibfield  {author} {\bibinfo {author} {\bibfnamefont {K.}~\bibnamefont {Momma}}\ and\ \bibinfo {author} {\bibfnamefont {F.}~\bibnamefont {Izumi}},\ }\href {https://doi.org/10.1107/S0021889811038970} {\bibfield  {journal} {\bibinfo  {journal} {J. Appl. Crystallogr.}\ }\textbf {\bibinfo {volume} {44}},\ \bibinfo {pages} {1272} (\bibinfo {year} {2011})}\BibitemShut {NoStop}%
\bibitem [{\citenamefont {Togo}\ \emph {et~al.}(2023)\citenamefont {Togo}, \citenamefont {Chaput}, \citenamefont {Tadano},\ and\ \citenamefont {Tanaka}}]{chapter3-6}%
  \BibitemOpen
  \bibfield  {author} {\bibinfo {author} {\bibfnamefont {A.}~\bibnamefont {Togo}}, \bibinfo {author} {\bibfnamefont {L.}~\bibnamefont {Chaput}}, \bibinfo {author} {\bibfnamefont {T.}~\bibnamefont {Tadano}},\ and\ \bibinfo {author} {\bibfnamefont {I.}~\bibnamefont {Tanaka}},\ }\href {https://doi.org/10.1088/1361-648X/acd831} {\bibfield  {journal} {\bibinfo  {journal} {J. Phys. Condens. Matter}\ }\textbf {\bibinfo {volume} {35}},\ \bibinfo {pages} {353001} (\bibinfo {year} {2023})}\BibitemShut {NoStop}%
\bibitem [{\citenamefont {Heyd}\ \emph {et~al.}(2003)\citenamefont {Heyd}, \citenamefont {Scuseria},\ and\ \citenamefont {Ernzerhof}}]{chapter1-35}%
  \BibitemOpen
  \bibfield  {author} {\bibinfo {author} {\bibfnamefont {J.}~\bibnamefont {Heyd}}, \bibinfo {author} {\bibfnamefont {G.~E.}\ \bibnamefont {Scuseria}},\ and\ \bibinfo {author} {\bibfnamefont {M.}~\bibnamefont {Ernzerhof}},\ }\href {https://doi.org/10.1063/1.1564060} {\bibfield  {journal} {\bibinfo  {journal} {J. Chem. Phys.}\ }\textbf {\bibinfo {volume} {118}},\ \bibinfo {pages} {8207} (\bibinfo {year} {2003})}\BibitemShut {NoStop}%
\bibitem [{\citenamefont {Wang}\ \emph {et~al.}(2021)\citenamefont {Wang}, \citenamefont {Xu}, \citenamefont {Liu}, \citenamefont {Tang},\ and\ \citenamefont {Geng}}]{chapter1-48}%
  \BibitemOpen
  \bibfield  {author} {\bibinfo {author} {\bibfnamefont {V.}~\bibnamefont {Wang}}, \bibinfo {author} {\bibfnamefont {N.}~\bibnamefont {Xu}}, \bibinfo {author} {\bibfnamefont {J.-C.}\ \bibnamefont {Liu}}, \bibinfo {author} {\bibfnamefont {G.}~\bibnamefont {Tang}},\ and\ \bibinfo {author} {\bibfnamefont {W.-T.}\ \bibnamefont {Geng}},\ }\href {https://doi.org/https://doi.org/10.1016/j.cpc.2021.108033} {\bibfield  {journal} {\bibinfo  {journal} {Comput. Phys. Commun.}\ }\textbf {\bibinfo {volume} {267}},\ \bibinfo {pages} {108033} (\bibinfo {year} {2021})}\BibitemShut {NoStop}%
\bibitem [{\citenamefont {Adhikari}\ and\ \citenamefont {Johari}(2023)}]{chapter2-38}%
  \BibitemOpen
  \bibfield  {author} {\bibinfo {author} {\bibfnamefont {S.}~\bibnamefont {Adhikari}}\ and\ \bibinfo {author} {\bibfnamefont {P.}~\bibnamefont {Johari}},\ }\href {https://doi.org/10.1103/PhysRevMaterials.7.075401} {\bibfield  {journal} {\bibinfo  {journal} {Phys. Rev. Mater.}\ }\textbf {\bibinfo {volume} {7}},\ \bibinfo {pages} {075401} (\bibinfo {year} {2023})}\BibitemShut {NoStop}%
\bibitem [{\citenamefont {Kangsabanik}\ \emph {et~al.}(2022)\citenamefont {Kangsabanik}, \citenamefont {Svendsen}, \citenamefont {Taghizadeh}, \citenamefont {Crovetto},\ and\ \citenamefont {Thygesen}}]{chapter3-35}%
  \BibitemOpen
  \bibfield  {author} {\bibinfo {author} {\bibfnamefont {J.}~\bibnamefont {Kangsabanik}}, \bibinfo {author} {\bibfnamefont {M.~K.}\ \bibnamefont {Svendsen}}, \bibinfo {author} {\bibfnamefont {A.}~\bibnamefont {Taghizadeh}}, \bibinfo {author} {\bibfnamefont {A.}~\bibnamefont {Crovetto}},\ and\ \bibinfo {author} {\bibfnamefont {K.~S.}\ \bibnamefont {Thygesen}},\ }\href {https://doi.org/10.1021/jacs.2c07567} {\bibfield  {journal} {\bibinfo  {journal} {J. Am. Chem. Soc.}\ }\textbf {\bibinfo {volume} {144}},\ \bibinfo {pages} {19872} (\bibinfo {year} {2022})},\ \bibinfo {note} {pMID: 36270007},\ \Eprint {https://arxiv.org/abs/https://doi.org/10.1021/jacs.2c07567} {https://doi.org/10.1021/jacs.2c07567} \BibitemShut {NoStop}%
\bibitem [{\citenamefont {Filip}\ \emph {et~al.}(2021)\citenamefont {Filip}, \citenamefont {Haber},\ and\ \citenamefont {Neaton}}]{chapter3-39}%
  \BibitemOpen
  \bibfield  {author} {\bibinfo {author} {\bibfnamefont {M.~R.}\ \bibnamefont {Filip}}, \bibinfo {author} {\bibfnamefont {J.~B.}\ \bibnamefont {Haber}},\ and\ \bibinfo {author} {\bibfnamefont {J.~B.}\ \bibnamefont {Neaton}},\ }\href {https://doi.org/10.1103/PhysRevLett.127.067401} {\bibfield  {journal} {\bibinfo  {journal} {Phys. Rev. Lett.}\ }\textbf {\bibinfo {volume} {127}},\ \bibinfo {pages} {067401} (\bibinfo {year} {2021})}\BibitemShut {NoStop}%
\bibitem [{\citenamefont {Hellwarth}\ and\ \citenamefont {Biaggio}(1999)}]{chapter2-22}%
  \BibitemOpen
  \bibfield  {author} {\bibinfo {author} {\bibfnamefont {R.~W.}\ \bibnamefont {Hellwarth}}\ and\ \bibinfo {author} {\bibfnamefont {I.}~\bibnamefont {Biaggio}},\ }\href {https://doi.org/10.1103/PhysRevB.60.299} {\bibfield  {journal} {\bibinfo  {journal} {Phys. Rev. B}\ }\textbf {\bibinfo {volume} {60}},\ \bibinfo {pages} {299} (\bibinfo {year} {1999})}\BibitemShut {NoStop}%
\bibitem [{\citenamefont {Adhikari}\ and\ \citenamefont {Johari}(2024{\natexlab{b}})}]{chapter5-18}%
  \BibitemOpen
  \bibfield  {author} {\bibinfo {author} {\bibfnamefont {S.}~\bibnamefont {Adhikari}}\ and\ \bibinfo {author} {\bibfnamefont {P.}~\bibnamefont {Johari}},\ }\href {https://doi.org/10.1103/PhysRevB.110.014101} {\bibfield  {journal} {\bibinfo  {journal} {Phys. Rev. B}\ }\textbf {\bibinfo {volume} {110}},\ \bibinfo {pages} {014101} (\bibinfo {year} {2024}{\natexlab{b}})}\BibitemShut {NoStop}%
\bibitem [{\citenamefont {Feynman}(1955)}]{chapter2-23}%
  \BibitemOpen
  \bibfield  {author} {\bibinfo {author} {\bibfnamefont {R.~P.}\ \bibnamefont {Feynman}},\ }\href {https://doi.org/10.1103/PhysRev.97.660} {\bibfield  {journal} {\bibinfo  {journal} {Phys. Rev.}\ }\textbf {\bibinfo {volume} {97}},\ \bibinfo {pages} {660} (\bibinfo {year} {1955})}\BibitemShut {NoStop}%
\bibitem [{\citenamefont {Xu}\ \emph {et~al.}(2018)\citenamefont {Xu}, \citenamefont {Liu}, \citenamefont {Wang}, \citenamefont {Liu},\ and\ \citenamefont {Huang}}]{chapter1-36}%
  \BibitemOpen
  \bibfield  {author} {\bibinfo {author} {\bibfnamefont {J.}~\bibnamefont {Xu}}, \bibinfo {author} {\bibfnamefont {J.-B.}\ \bibnamefont {Liu}}, \bibinfo {author} {\bibfnamefont {J.}~\bibnamefont {Wang}}, \bibinfo {author} {\bibfnamefont {B.-X.}\ \bibnamefont {Liu}},\ and\ \bibinfo {author} {\bibfnamefont {B.}~\bibnamefont {Huang}},\ }\href {https://doi.org/https://doi.org/10.1002/adfm.201800332} {\bibfield  {journal} {\bibinfo  {journal} {Adv. Funct. Mater.}\ }\textbf {\bibinfo {volume} {28}},\ \bibinfo {pages} {1800332} (\bibinfo {year} {2018})}\BibitemShut {NoStop}%
\bibitem [{\citenamefont {Li}\ \emph {et~al.}(2008)\citenamefont {Li}, \citenamefont {Lu}, \citenamefont {Ding}, \citenamefont {Feng}, \citenamefont {Gao},\ and\ \citenamefont {Guo}}]{chapter1-37}%
  \BibitemOpen
  \bibfield  {author} {\bibinfo {author} {\bibfnamefont {C.}~\bibnamefont {Li}}, \bibinfo {author} {\bibfnamefont {X.}~\bibnamefont {Lu}}, \bibinfo {author} {\bibfnamefont {W.}~\bibnamefont {Ding}}, \bibinfo {author} {\bibfnamefont {L.}~\bibnamefont {Feng}}, \bibinfo {author} {\bibfnamefont {Y.}~\bibnamefont {Gao}},\ and\ \bibinfo {author} {\bibfnamefont {Z.}~\bibnamefont {Guo}},\ }\href {https://doi.org/10.1107/S0108768108032734} {\bibfield  {journal} {\bibinfo  {journal} {Acta Crystallogr.}\ }\textbf {\bibinfo {volume} {64}},\ \bibinfo {pages} {702} (\bibinfo {year} {2008})}\BibitemShut {NoStop}%
\bibitem [{\citenamefont {Volonakis}\ \emph {et~al.}(2017)\citenamefont {Volonakis}, \citenamefont {Haghighirad}, \citenamefont {Milot}, \citenamefont {Sio}, \citenamefont {Filip}, \citenamefont {Wenger}, \citenamefont {Johnston}, \citenamefont {Herz}, \citenamefont {Snaith},\ and\ \citenamefont {Giustino}}]{chapter1-38}%
  \BibitemOpen
  \bibfield  {author} {\bibinfo {author} {\bibfnamefont {G.}~\bibnamefont {Volonakis}}, \bibinfo {author} {\bibfnamefont {A.~A.}\ \bibnamefont {Haghighirad}}, \bibinfo {author} {\bibfnamefont {R.~L.}\ \bibnamefont {Milot}}, \bibinfo {author} {\bibfnamefont {W.~H.}\ \bibnamefont {Sio}}, \bibinfo {author} {\bibfnamefont {M.~R.}\ \bibnamefont {Filip}}, \bibinfo {author} {\bibfnamefont {B.}~\bibnamefont {Wenger}}, \bibinfo {author} {\bibfnamefont {M.~B.}\ \bibnamefont {Johnston}}, \bibinfo {author} {\bibfnamefont {L.~M.}\ \bibnamefont {Herz}}, \bibinfo {author} {\bibfnamefont {H.~J.}\ \bibnamefont {Snaith}},\ and\ \bibinfo {author} {\bibfnamefont {F.}~\bibnamefont {Giustino}},\ }\href {https://doi.org/10.1021/acs.jpclett.6b02682} {\bibfield  {journal} {\bibinfo  {journal} {J. Phys. Chem. Lett.}\ }\textbf {\bibinfo {volume} {8}},\ \bibinfo {pages} {772} (\bibinfo {year} {2017})}\BibitemShut {NoStop}%
\bibitem [{\citenamefont {Mouhat}\ and\ \citenamefont {Coudert}(2014)}]{chapter1-47}%
  \BibitemOpen
  \bibfield  {author} {\bibinfo {author} {\bibfnamefont {F.}~\bibnamefont {Mouhat}}\ and\ \bibinfo {author} {\bibfnamefont {F.-X.}\ \bibnamefont {Coudert}},\ }\href {https://doi.org/10.1103/PhysRevB.90.224104} {\bibfield  {journal} {\bibinfo  {journal} {Phys. Rev. B}\ }\textbf {\bibinfo {volume} {90}},\ \bibinfo {pages} {224104} (\bibinfo {year} {2014})}\BibitemShut {NoStop}%
\bibitem [{\citenamefont {Wu}\ \emph {et~al.}(2007)\citenamefont {Wu}, \citenamefont {Zhao}, \citenamefont {Xiang}, \citenamefont {Hao}, \citenamefont {Liu},\ and\ \citenamefont {Meng}}]{chapter1-49}%
  \BibitemOpen
  \bibfield  {author} {\bibinfo {author} {\bibfnamefont {Z.-j.}\ \bibnamefont {Wu}}, \bibinfo {author} {\bibfnamefont {E.-j.}\ \bibnamefont {Zhao}}, \bibinfo {author} {\bibfnamefont {H.-p.}\ \bibnamefont {Xiang}}, \bibinfo {author} {\bibfnamefont {X.-f.}\ \bibnamefont {Hao}}, \bibinfo {author} {\bibfnamefont {X.-j.}\ \bibnamefont {Liu}},\ and\ \bibinfo {author} {\bibfnamefont {J.}~\bibnamefont {Meng}},\ }\href {https://doi.org/10.1103/PhysRevB.76.054115} {\bibfield  {journal} {\bibinfo  {journal} {Phys. Rev. B}\ }\textbf {\bibinfo {volume} {76}},\ \bibinfo {pages} {054115} (\bibinfo {year} {2007})}\BibitemShut {NoStop}%
\bibitem [{\citenamefont {Hill}(1952)}]{chapter1-50}%
  \BibitemOpen
  \bibfield  {author} {\bibinfo {author} {\bibfnamefont {R.}~\bibnamefont {Hill}},\ }\href {https://doi.org/10.1088/0370-1298/65/5/307} {\bibfield  {journal} {\bibinfo  {journal} {Proc. Phys. Soc. A}\ }\textbf {\bibinfo {volume} {65}},\ \bibinfo {pages} {349} (\bibinfo {year} {1952})}\BibitemShut {NoStop}%
\bibitem [{\citenamefont {Pugh}(1954)}]{chapter1-51}%
  \BibitemOpen
  \bibfield  {author} {\bibinfo {author} {\bibfnamefont {S.}~\bibnamefont {Pugh}},\ }\href {https://doi.org/10.1080/14786440808520496} {\bibfield  {journal} {\bibinfo  {journal} {London, Edinburgh Dublin Philos. Mag. J. Sci.}\ }\textbf {\bibinfo {volume} {45}},\ \bibinfo {pages} {823} (\bibinfo {year} {1954})}\BibitemShut {NoStop}%
\bibitem [{\citenamefont {Wei}\ \emph {et~al.}(2020)\citenamefont {Wei}, \citenamefont {Hui}, \citenamefont {Perera}, \citenamefont {Sheng}, \citenamefont {Watson}, \citenamefont {Sun}, \citenamefont {Jia}, \citenamefont {Zhang},\ and\ \citenamefont {Zeng}}]{chapter6-2}%
  \BibitemOpen
  \bibfield  {author} {\bibinfo {author} {\bibfnamefont {X.}~\bibnamefont {Wei}}, \bibinfo {author} {\bibfnamefont {H.}~\bibnamefont {Hui}}, \bibinfo {author} {\bibfnamefont {S.}~\bibnamefont {Perera}}, \bibinfo {author} {\bibfnamefont {A.}~\bibnamefont {Sheng}}, \bibinfo {author} {\bibfnamefont {D.~F.}\ \bibnamefont {Watson}}, \bibinfo {author} {\bibfnamefont {Y.-Y.}\ \bibnamefont {Sun}}, \bibinfo {author} {\bibfnamefont {Q.}~\bibnamefont {Jia}}, \bibinfo {author} {\bibfnamefont {S.}~\bibnamefont {Zhang}},\ and\ \bibinfo {author} {\bibfnamefont {H.}~\bibnamefont {Zeng}},\ }\href {https://doi.org/10.1021/acsomega.0c00740} {\bibfield  {journal} {\bibinfo  {journal} {ACS Omega}\ }\textbf {\bibinfo {volume} {5}},\ \bibinfo {pages} {18579} (\bibinfo {year} {2020})},\ \Eprint {https://arxiv.org/abs/https://doi.org/10.1021/acsomega.0c00740} {https://doi.org/10.1021/acsomega.0c00740} \BibitemShut {NoStop}%
\bibitem [{\citenamefont {Bokdam}\ \emph {et~al.}(2016)\citenamefont {Bokdam}, \citenamefont {Sander}, \citenamefont {Stroppa}, \citenamefont {Picozzi}, \citenamefont {Sarma}, \citenamefont {Franchini},\ and\ \citenamefont {Kresse}}]{chapter1-65}%
  \BibitemOpen
  \bibfield  {author} {\bibinfo {author} {\bibfnamefont {M.}~\bibnamefont {Bokdam}}, \bibinfo {author} {\bibfnamefont {T.}~\bibnamefont {Sander}}, \bibinfo {author} {\bibfnamefont {A.}~\bibnamefont {Stroppa}}, \bibinfo {author} {\bibfnamefont {S.}~\bibnamefont {Picozzi}}, \bibinfo {author} {\bibfnamefont {D.~D.}\ \bibnamefont {Sarma}}, \bibinfo {author} {\bibfnamefont {C.}~\bibnamefont {Franchini}},\ and\ \bibinfo {author} {\bibfnamefont {G.}~\bibnamefont {Kresse}},\ }\href {https://doi.org/10.1038/srep28618} {\bibfield  {journal} {\bibinfo  {journal} {Sci. Rep.}\ }\textbf {\bibinfo {volume} {6}},\ \bibinfo {pages} {28618} (\bibinfo {year} {2016})}\BibitemShut {NoStop}%
\bibitem [{\citenamefont {Spataru}\ and\ \citenamefont {Leonard}(2013)}]{chapter1-66}%
  \BibitemOpen
  \bibfield  {author} {\bibinfo {author} {\bibfnamefont {C.~D.}\ \bibnamefont {Spataru}}\ and\ \bibinfo {author} {\bibfnamefont {F.}~\bibnamefont {Leonard}},\ }\href {https://doi.org/https://doi.org/10.1016/j.chemphys.2012.08.021} {\bibfield  {journal} {\bibinfo  {journal} {Chem. Phys.}\ }\textbf {\bibinfo {volume} {413}},\ \bibinfo {pages} {81} (\bibinfo {year} {2013})}\BibitemShut {NoStop}%
\bibitem [{\citenamefont {Herz}(2017)}]{chapter2-21}%
  \BibitemOpen
  \bibfield  {author} {\bibinfo {author} {\bibfnamefont {L.~M.}\ \bibnamefont {Herz}},\ }\href {https://doi.org/10.1021/acsenergylett.7b00276} {\bibfield  {journal} {\bibinfo  {journal} {ACS Energy Lett.}\ }\textbf {\bibinfo {volume} {2}},\ \bibinfo {pages} {1539} (\bibinfo {year} {2017})},\ \Eprint {https://arxiv.org/abs/https://doi.org/10.1021/acsenergylett.7b00276} {https://doi.org/10.1021/acsenergylett.7b00276} \BibitemShut {NoStop}%
\bibitem [{\citenamefont {Fr\"ohlich}(1954)}]{chapter2-51}%
  \BibitemOpen
  \bibfield  {author} {\bibinfo {author} {\bibfnamefont {H.}~\bibnamefont {Fr\"ohlich}},\ }\href {https://doi.org/10.1080/00018735400101213} {\bibfield  {journal} {\bibinfo  {journal} {Adv. Phys.}\ }\textbf {\bibinfo {volume} {3}},\ \bibinfo {pages} {325} (\bibinfo {year} {1954})},\ \Eprint {https://arxiv.org/abs/https://doi.org/10.1080/00018735400101213} {https://doi.org/10.1080/00018735400101213} \BibitemShut {NoStop}%
\bibitem [{\citenamefont {Yu}\ and\ \citenamefont {Zunger}(2012)}]{chapter3-30}%
  \BibitemOpen
  \bibfield  {author} {\bibinfo {author} {\bibfnamefont {L.}~\bibnamefont {Yu}}\ and\ \bibinfo {author} {\bibfnamefont {A.}~\bibnamefont {Zunger}},\ }\href {https://doi.org/10.1103/PhysRevLett.108.068701} {\bibfield  {journal} {\bibinfo  {journal} {Phys. Rev. Lett.}\ }\textbf {\bibinfo {volume} {108}},\ \bibinfo {pages} {068701} (\bibinfo {year} {2012})}\BibitemShut {NoStop}%
\bibitem [{\citenamefont {Shockley}\ and\ \citenamefont {Queisser}(2004)}]{chapter3-31}%
  \BibitemOpen
  \bibfield  {author} {\bibinfo {author} {\bibfnamefont {W.}~\bibnamefont {Shockley}}\ and\ \bibinfo {author} {\bibfnamefont {H.~J.}\ \bibnamefont {Queisser}},\ }\href {https://doi.org/10.1063/1.1736034} {\bibfield  {journal} {\bibinfo  {journal} {J. Appl. Phys.}\ }\textbf {\bibinfo {volume} {32}},\ \bibinfo {pages} {510} (\bibinfo {year} {2004})},\ \Eprint {https://arxiv.org/abs/https://pubs.aip.org/aip/jap/article-pdf/32/3/510/10548356/510\_1\_online.pdf} {https://pubs.aip.org/aip/jap/article-pdf/32/3/510/10548356/510\_1\_online.pdf} \BibitemShut {NoStop}%
\bibitem [{\citenamefont {Dias}\ \emph {et~al.}(2021)\citenamefont {Dias}, \citenamefont {Lima},\ and\ \citenamefont {Da~Silva}}]{chapter3-33}%
  \BibitemOpen
  \bibfield  {author} {\bibinfo {author} {\bibfnamefont {A.~C.}\ \bibnamefont {Dias}}, \bibinfo {author} {\bibfnamefont {M.~P.}\ \bibnamefont {Lima}},\ and\ \bibinfo {author} {\bibfnamefont {J.~L.~F.}\ \bibnamefont {Da~Silva}},\ }\href {https://doi.org/10.1021/acs.jpcc.1c05245} {\bibfield  {journal} {\bibinfo  {journal} {J. Phys. Chem. C}\ }\textbf {\bibinfo {volume} {125}},\ \bibinfo {pages} {19142} (\bibinfo {year} {2021})},\ \Eprint {https://arxiv.org/abs/https://doi.org/10.1021/acs.jpcc.1c05245} {https://doi.org/10.1021/acs.jpcc.1c05245} \BibitemShut {NoStop}%
\bibitem [{\citenamefont {Qian}\ \emph {et~al.}(2016)\citenamefont {Qian}, \citenamefont {Xu},\ and\ \citenamefont {Tian}}]{chapter3-34}%
  \BibitemOpen
  \bibfield  {author} {\bibinfo {author} {\bibfnamefont {J.}~\bibnamefont {Qian}}, \bibinfo {author} {\bibfnamefont {B.}~\bibnamefont {Xu}},\ and\ \bibinfo {author} {\bibfnamefont {W.}~\bibnamefont {Tian}},\ }\href {https://doi.org/https://doi.org/10.1016/j.orgel.2016.05.046} {\bibfield  {journal} {\bibinfo  {journal} {Org. Electron.}\ }\textbf {\bibinfo {volume} {37}},\ \bibinfo {pages} {61} (\bibinfo {year} {2016})}\BibitemShut {NoStop}%
\bibitem [{\citenamefont {Kangsabanik}\ \emph {et~al.}(2018)\citenamefont {Kangsabanik}, \citenamefont {Sugathan}, \citenamefont {Yadav}, \citenamefont {Yella},\ and\ \citenamefont {Alam}}]{chapter3-32}%
  \BibitemOpen
  \bibfield  {author} {\bibinfo {author} {\bibfnamefont {J.}~\bibnamefont {Kangsabanik}}, \bibinfo {author} {\bibfnamefont {V.}~\bibnamefont {Sugathan}}, \bibinfo {author} {\bibfnamefont {A.}~\bibnamefont {Yadav}}, \bibinfo {author} {\bibfnamefont {A.}~\bibnamefont {Yella}},\ and\ \bibinfo {author} {\bibfnamefont {A.}~\bibnamefont {Alam}},\ }\href {https://doi.org/10.1103/PhysRevMaterials.2.055401} {\bibfield  {journal} {\bibinfo  {journal} {Phys. Rev. Mater.}\ }\textbf {\bibinfo {volume} {2}},\ \bibinfo {pages} {055401} (\bibinfo {year} {2018})}\BibitemShut {NoStop}%
\end{thebibliography}%

\end{document}